\documentclass[usenatbib,usegraphicx]{mn2e}
\usepackage{bm}
\usepackage{morefloats}
\usepackage{hyperref}
\usepackage{tabularx}
\usepackage{graphicx}
\usepackage{multirow}
\usepackage{morefloats}
\usepackage{pdflscape}

\title[Structural parameters of 112 globular clusters]
{A catalogue of masses, structural parameters and velocity dispersion profiles of 112 Milky Way globular clusters}
\author[Baumgardt \& Hilker]{H. Baumgardt$^{1}$\thanks{E-mail:
h.baumgardt@uq.edu.au}, M. Hilker$^{2}$\\
$^{1}$ School of Mathematics and Physics, The University of Queensland, St. Lucia, QLD 4072, Australia \\
$^{2}$ European Southern Observatory, Karl-Schwarzschild-Str. 2, 85748 Garching, Germany\\
}

\begin{document}

\date{Accepted 2017 xx xx. Received 2017 xx xx; in original form 2017 xx xx}

\pagerange{\pageref{firstpage}--\pageref{lastpage}} \pubyear{201x}

\maketitle

\label{firstpage}

\begin{abstract}
We have determined masses, stellar mass functions and structural parameters of 112 Milky Way globular clusters by
fitting a large set of $N$-body simulations to their velocity dispersion and surface density profiles.
The velocity dispersion profiles were calculated based on a combination of more than 15,000 high-precision radial 
velocities which we derived from archival ESO/VLT and Keck spectra together with $\sim 20,000$ published radial velocities
from the literature. Our fits also include the stellar mass functions of the globular clusters, which are available 
for 47 clusters in our sample, allowing us to self-consistently take the effects of mass segregation
and ongoing cluster dissolution into account. We confirm the strong correlation between the 
global mass functions of globular clusters and their relaxation times recently found by \citet{sollimabaumgardt2017}.
We also find a correlation 
of the escape velocity from the centre of a globular cluster and the fraction of first generation stars (FG)
in the cluster recently derived for 57 globular clusters by
\citet{miloneetal2017}, but no correlation between the FG star fraction and the global mass function
of a globular cluster. This could indicate that the ability of a globular cluster
to keep the wind ejecta from the polluting star(s) is the crucial parameter determining the presence and
fraction of second generation stars and not its later dynamical mass loss.
\end{abstract}


\section{Introduction} \label{sec:intro}

Globular clusters are excellent laboratories to study star formation and the early evolution of galaxies
since they contain large samples of equidistant stars that have coeval ages (at least to within a few tens of Myr) 
and similar chemical abundance patterns (at least for heavy elements).
Measuring the properties of stars in globular clusters therefore allows to accurately determine many 
of the fundamental parameters of globular clusters like distances, ages, metallicities, sizes and masses.
In addition, their high stellar densities make them unique environments for the creation of exotic stars like blue 
stragglers \citep{bailyn1995,davies2004}, low-mass X-ray binaries \citep{verbunt1993,pooleyetal2003} and millisecond 
pulsars \citep{manchesteretal1991}. Globular clusters are also among the prime environments for the
creation of black hole binaries that are tight enough so they can merge through the emission of gravitational 
waves within a Hubble time \citep{banerjeeetal2010,downingetal2011,rodriguezetal2016a,askaretal2017}.
They are also interesting from a theoretical point of view since they allow to study the interplay between
stellar evolution, binary evolution and stellar dynamics.

An accurate understanding of the current state and evolutionary history of a globular cluster requires a detailed knowledge of its
internal mass distribution as well as the current stellar mass function. In recent years, 
information on the velocity dispersion profiles has become available through large surveys using either multi-object spectrographs on 4m and 8m class telescopes 
\citep[e.g.][]{laneetal2011,kimmigetal2015,lardoetal2015,kamannetal2018} or proper motions of stars using HST \citep{watkinsetal2015a}.
In addition, HST photometry has allowed to determine the stellar mass functions of many globular clusters from the
tip of the red giant branch down to almost the hydrogen burning limit \citep[e.g.][]{demarchietal2007,paustetal2010,webbleigh2015,sollimabaumgardt2017}.
At the same time, analytic models like King-Michie models \citep[e.g.][]{sollimaetal2012} or models based on the solution of
lowered isothermal distribution functions like the {\tt LIMEPY} models \citep{gieleszocchi2015}
have become sophisticated enough to model the internal mass distribution of a globular cluster including mass-segregation.
Furthermore, progress in the speed of computers as well as increasing sophistication of the computer codes has allowed to perform simulations
of globular clusters with up to $10^6$ stars through either direct $N$-body \citep{heggie2014,wangetal2016} or Monte Carlo simulations \citep[e.g.][]{gierszheggie2011,askaretal2017}, meaning that a
detailed comparison of observations and simulations for individual globular clusters has become possible \citep[e.g.][]{zonoozietal2011}.
While analytic models are flexible and fast, direct simulation methods are naturally self-consistent and offer the opportunity to put constraints 
on the initial conditions of Milky Way globular clusters.

\citet{baumgardt2017} have recently derived total masses and mass-to-light ratios of 50 Galactic globular clusters based on a comparison
of their velocity dispersion and surface density profiles with the results of a large set of $N$-body simulations. In the current paper we
improve their modeling by including the stellar mass functions of globular clusters 
in our modeling. We do this by calculating new sets of models that have mass functions that are depleted in low-mass stars compared
to a canonical Kroupa (2001) mass function. We also significantly improve the velocity dispersion profiles calculated by \citet{baumgardt2017} based on 
individual stellar radial velocities
by additional radial velocities from unpublished spectra from the ESO and Keck science archives.  Our paper is organized as follows:
In section 2 we describe the observational data used in this paper and the reduction of the spectra. In section 3 we present the
new grid of $N$-body simulations that we have calculated and section 4 presents our results. We draw our conclusions in section~5. 

\section{Observational data}

\subsection{Radial velocities}

The radial velocities used in this work were derived from mainly two sources: We first searched the ESO and Keck Science archives for unpublished spectra of stars in globular clusters. For ESO/VLT spectra,
we searched the ESO Science Archive for reduced, high-resolution {\tt FLAMES}, {\tt UVES}, {\tt X-Shooter} and {\tt FEROS} spectra of stars within 15 arcmin around the center
of each globular cluster. If the spectra were not already in a heliocentric reference frame, we first applied a heliocentric correction to them using the 
{\tt bcvcorr} routine from the {\tt RVSAO} software package \citep{kurtzmink1998}.
{\tt FLAMES} spectra were then sky subtracted with the help of the {\tt Skycorr} package \citep{nolletal2014}. In order to perform the sky subtraction, we used the median 
of the 8 associated sky fibers as the sky reference spectrum for each stellar spectrum. We then
co-added individual spectra taken within 30 days of each other using the IRAF\footnote{IRAF is distributed by the National Optical Astronomy Observatories, which are operated by the Association of Universities for Research
in Astronomy, Inc., under cooperative agreement with the National Science Foundation.} {\tt scombine} task and
determined stellar radial velocities with the help of the IRAF {\tt fxcor} task, using as templates the spectra of cool giant
stars of a metallicity that is comparable to the cluster metallicity given in \citet{harris1996}. We created the template spectra with the stellar synthesis program {\tt SPECTRUM} \citep{graycorbally1994} using 
{\tt ATLAS9} stellar model atmospheres \citep{castellikurucz2004}.
For each cluster we created the template spectrum from the theoretical atmosphere models that were closest in metallicity to the studied clusters
and used the same spectral resolution as the observed spectra.

For a few clusters we also determined radial velocities from 
archival ESO/VLT {\tt FORS2} spectra that include the Calcium triplet lines. In order to derive radial velocities from {\tt FORS2} spectra, we reduced the raw data with the help
of the ESO Reflex pipeline \citep{freudlingetal2013}. We then again ran the IRAF {\tt fxcor} task and applied a telluric correction to the spectra similar to the analysis of the Keck
{\tt DEIMOS} spectra described below.

We also derived radial velocities from unpublished Keck {\tt DEIMOS}, {\tt HIRES}, and {\tt NIRSPEC} spectra available from the Keck Observatory archive. 
The {\tt DEIMOS} spectra were reduced with the {\tt DEEP2} data reduction pipeline developed by the {\tt DEEP2} survey team 
\citep{cooperetal2012, newmanetal2013}, while for {\tt HIRES} and {\tt NIRSPEC} data we used the reduced spectra already available in the Keck Observatory archive.
We then used {\tt fxcor} to derive the stellar radial velocities from the spectra. 
{\tt HIRES} and {\tt DEIMOS} spectra were again cross-correlated against synthetic template spectra created with the {\tt SPECTRUM} synthesis program, 
while the {\tt NIRSPEC} spectra
were cross-correlated against the infrared spectrum of the K giant star 10 Leonis available from the {\tt CRIRES} spectral library of \citet{nichollsetal2017}.
In order to correct for residual systematic errors in the absolute wavelength calibration of the {\tt DEIMOS} spectra, we cross-correlated them against a telluric template
spectrum that was kindly provided to us by Tony Sohn and Emily Cunningham. Since the telluric lines should be at zero radial velocity, systematic wavelength calibration 
errors can be corrected from the radial velocity of these lines. Final radial velocities for the {\tt DEIMOS} spectra were then calculated according to $v_r=v_{obs}-v_{tel}-v_{hel}$, where $v_{obs}$ 
is the radial velocity derived from the stellar template, $v_{tel}$ the radial velocity from the telluric spectrum and $v_{hel}$ the heliocentric
correction. 

In order to improve the accuracy of the stellar positions for all GCs, we cross-correlated the stellar positions given in the FITS file headers 
of the individual spectra against the stellar positions in the {\tt 2MASS} catalogue \citep{skrutskieetal2006}, 
supplemented in a few cases by HST/ACS data or other catalogues. The stellar positions given in the FITS file headers were replaced
whenever a matching position within 2 arcsec was found in {\tt 2MASS} or one of the other catalogues.

In total we could derive about 15,000 radial velocities from unpublished ESO/VLT and Keck spectra for stars in about 90 globular clusters.
Tables~\ref{indveltabstart} to \ref{indveltabend} (made available in their full extend online) list the individual stellar radial velocities 
that we derived from ESO {\tt FLAMES} and {\tt UVES} observations done before 2014. The membership probabilities $P_i$ in Tables~\ref{indveltabstart} to \ref{indveltabend} were calculated
according to $P_i=1.0-\mbox{erf}(\sqrt{\chi_i^2}/0.5)$ where erf is the error function and $\chi_i^2$ is the error and velocity dispersion weighted difference between the individual stellar radial velocity $v_i$ and the mean cluster velocity $<\!v\!>$ calculated according to
\begin{equation}
\chi_i^2 = e^{-\frac{1}{2}\frac{\left(v_i-<\!v\!>\right)^2}{\epsilon_i^2+\sigma_r^2}}  \;\; ,
\end{equation}
Here $\epsilon_i$ is the error of the stellar velocity and $\sigma_r$ is the expected velocity dispersion at the projected distance of
the star calculated from the best-fitting $N$-body model. 

We supplemented the VLT and Keck data by published radial velocities from the literature.  
Our main source of published literature data is the recent compilation by \citet{baumgardt2017}. Additional 
literature data used in this work is given in Table~\ref{tab:litdata}. We include the velocity dispersion profiles
recently published by \citet{kamannetal2018} from a {\tt MUSE} survey of the centers of 25 globular clusters. In order to allow for easy comparison
with the other available radial velocity data, which is mainly restricted to giant stars, we restrict
ourselves to clusters with a high effective mass in Table~5 of \citet{kamannetal2018}, i.e. clusters where the {\tt MUSE} velocity dispersion profiles
are dominated by massive turn-off and giant stars. We also replaced the APOGEE DR13 radial velocities 
used in \citet{baumgardt2017} with the radial velocities published in the APOGEE DR14 data release \citep{abolfathietal2017}.

Our final sample consists of 42,000 radial velocity measurements of about 35,000 individual stars in 109 globular
clusters. The median uncertainty of an individual measurement is about 0.5 km/sec and 90\% of our stars have velocity errors of less than
2 km/sec. The errors should therefore be small enough 
to reliably derive the velocity dispersion profiles of most globular clusters, except for the lowest mass clusters in which the internal 
velocity dispersion is less than 1 km/sec.

\subsection{Radial velocity dispersion profiles}

\begin{figure}
\begin{center}
\includegraphics[width=\columnwidth]{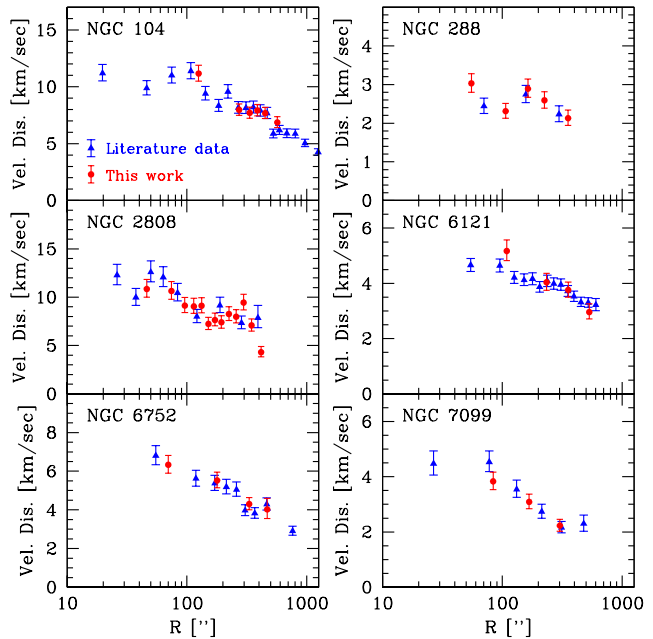}
\end{center}
\caption{Comparison of the velocity dispersion profile derived from literature radial velocities (blue triangles) vs. the velocity dispersion profile
derived from radial velocities determined in this work (red circles) for six clusters which have more than 300 radial velocities
in both data sets. It can be seen that the velocity dispersion profile determined from our radial velocities is 
in excellent agreement with the one based on literature radial velocities.}
\label{comparison}
\end{figure}

In order to derive velocity dispersion profiles from the individual stellar radial velocities, we cross-correlated the radial velocities from the different data sets
against each other to bring them to a common mean radial velocity. The necessary radial velocity shifts were usually less than 1 km/sec. We then merged the individual data
sets to create a master catalogue 
for each globular cluster. Multiple measurements of individual stars were averaged and we performed a $\chi^2$ test to evaluate whether the measured individual
radial velocities were compatible with a constant radial velocity. 
Stars where the individual measurements had a less than 5\% probability to be compatible
with a constant radial velocity were removed before 
the velocity dispersion profile was calculated. The velocity dispersion profile of each cluster was then determined using a maximum-likelihood approach.
Non-members were removed iteratively during the calculation of the velocity dispersion profiles. More details on the way the velocity dispersion 
profiles were calculated can be found in \citet{baumgardt2017}. Table \ref{veldistab} presents the velocity dispersion profiles of all 
clusters calculated from the individual stellar radial velocities.\footnote{The radial velocity dispersion profiles including possible updates if new data has become available since the
publication of this paper can also be downloaded from
\href{https://people.smp.uq.edu.au/HolgerBaumgardt/globular/}{https://people.smp.uq.edu.au/HolgerBaumgardt/globular/}}
Fig.\ \ref{comparison} shows a comparison of the velocity dispersion profiles
calculated from literature data alone (blue triangles) vs. the velocity dispersion profiles that we obtain using only radial velocities
determined from VLT/Keck spectra for the six globular clusters which have the largest number of stars in both data sets. It can be seen that the profiles are in excellent agreement with 
each other.

NGC~2298 and Pal~5 have velocity dispersion profiles which increase towards the outer parts, which could
be due to the ongoing tidal disruption of these clusters \citep{odenkirchenetal2002, balbinotetal2011}. We therefore neglected the outermost data point of their
velocity dispersion profiles in our fits. \citet{blechaetal2004} and \citet{bardfordetal2011} found that the velocity dispersion profile of Pal~13 could be
inflated by binaries. In order to reduce the
effect of these binaries, we only take stars with more than one radial velocity measurement into account when calculating the velocity dispersion of this cluster. This
reduces the velocity dispersion by about 50\% compared to the case when we use all stars. However the resulting $M/L$ ratio is still significantly larger
than what we find for other globular clusters, indicating that undetected binaries might still be present in the Pal~13 sample. Undetected binaries might also be present
in a few other low-mass and low density clusters like Arp~2 or Ter~3 which also have $M/L$ values significantly higher than the rest of the clusters.

\subsection{Stellar mass functions and other cluster data}

We took most stellar mass functions from \cite{sollimabaumgardt2017} who determined stellar mass functions in four annuli inside 
a projected radius of $r<1.6'$ around the centers of 35 globular clusters using the HST/ACS photometry published by the {\it Globular Cluster ACS Treasury Project} 
\citep{sarajedinietal2007}. For most clusters these mass functions cover the area inside the half-light radius and the mass range between
0.2 M$_\odot< m < 0.8$~M$_\odot$, although for massive and concentrated clusters, low mass stars are too faint to be observed in the 
cluster centres. In addition, we also searched the literature for additional deep, completeness corrected HST/ACS and WFPC2 
photometry of the luminosity function of main sequence stars in globular clusters. This data often complements the ACS Treasury Project data
at larger radii, giving a more complete spatial coverage of the stellar mass function profile for a particular cluster. We determined 
the stellar mass functions
from the luminosity functions by fitting either Dartmouth \citep{dotteretal2008} or PARSEC isochrones \citep{bressanetal2012} to 
the HST color-magnitude diagrams. The additional photometric
data is listed in Table~\ref{tab:litdata}. In total we could determine stellar mass functions of 47 globular clusters, i.e. 
roughly half of all globular clusters in our sample. 

Together with the stellar radial velocity dispersion profiles, we also fitted the proper motion dispersion profiles published by \citet{watkinsetal2015a}. We used
their combined 1D profiles which are averaged over the radial and tangential component of the proper motions. For clusters with
available proper motions, we varied the cluster distances until we obtained the best agreement (lowest combined $\chi^2$) in the simultaneous fit of radial velocity and proper motion
dispersion profile. Clusters where distances were fitted are indicated in column 5 of Table~2.
The surface density profiles were mostly taken from \citet{trageretal1995} and, if available, from \citet{noyolagebhardt2006}. Clusters where we used 
other surface density profiles are listed in Table~\ref{tab:litdata}. Apparent $V$-band magnitudes and their
errors were calculated by taking the average of the apparent magnitudes given in \citet{harris1996}, \citet{mclaughlinvandermarel2005}, 
\citet{dalessandroetal2012} and the integrated magnitudes determined in this work from the fit of our models to the surface brightness profiles.
For the clusters in common with \citet{baumgardt2017} we took the ages from their paper. For the other clusters we searched the literature for
age determinations. If no age could be found for a particular cluster we assumed an age of 12 Gyr.

\begin{table*}
\caption{Power-law mass function slopes $N(m) \sim m^{\alpha}$ and mass limits used to set-up the $N$-body models.}
\begin{tabularx}{\textwidth}{cccccccccccccccc}
\hline
\multirow{2}{*}{Model} & $m_{Low}$ & $m_{Up}$ & \multirow{2}{*}{$\alpha$} & $m_{Low}$ & $m_{Up}$ & \multirow{2}{*}{$\alpha$} & $m_{Low}$ & $m_{Up}$ & \multirow{2}{*}{$\alpha$} & $m_{Low}$ & $m_{Up}$ & \multirow{2}{*}{$\alpha$} & $m_{Low}$ & $m_{Up}$ & \multirow{2}{*}{$\alpha$}\\
 & [M$_\odot$] & [M$_\odot$] & & [M$_\odot$] & [M$_\odot$] & & [M$_\odot$] & [M$_\odot$] & & [M$_\odot$] & [M$_\odot$] & & [M$_\odot$] & [M$_\odot$] &\\
\hline
1 & 0.10 & 0.20 & -1.35 & 0.20 & 0.50 & -1.35 & 0.50 & 0.80 & -2.35 & 0.80 & 1.0 & -2.35  & 1.00 & 15.0 & -2.35\\
2 & 0.10 & 0.20 & -1.05 & 0.20 & 0.50 & -0.80 & 0.50 & 0.80 & -1.70 & 0.80 & 1.0 & -2.20  & 1.00 & 15.0 & -2.20\\
3 & 0.10 & 0.20 & -0.85 & 0.20 & 0.50 & -0.30 & 0.50 & 0.80 & -1.05 & 0.80 & 1.0 & -2.20  & 1.00 & 15.0 & -2.00\\
4 & 0.10 & 0.20 & -0.60 & 0.20 & 0.50 & $\,\,$0.15 & 0.50 & 0.80 & -0.40 & 0.80 & 1.0 & -1.80  & 1.00 & 15.0 & -1.80\\
5 & 0.10 & 0.20 & -0.40 & 0.20 & 0.50 & $\,\,$0.55 & 0.50 & 0.80 &  $\,\,$0.30 & 0.80 & 1.0 & -3.00  & 1.00 & 15.0 & -1.60\\
\hline
\end{tabularx}
\label{tab:mfrange}
\end{table*}

\section{$N$-body models}

We determined cluster masses and structural parameters by comparing the observed velocity dispersion, surface density and
stellar mass function profiles against a grid of about 1200 $N$-body simulations. The details of the $N$-body simulations and the basic strategy to 
determine the best-fitting model are the same as described in \citet{baumgardt2017} and we refer the reader to this paper
for a detailed description of the modeling. In short, we ran $N$-body simulations of isolated star clusters, each containing $N=100,000$ 
stars initially using the GPU-enabled version of the collisional $N$-body code NBODY6 \citep{aarseth1999,nitadoriaarseth2012}. The simulated
clusters followed \citet{king1962} density profiles initially. The initial concentrations were varied between $0.2 \le c \le 2.5$ and the
initial radii were varied between $2 \le r_h \le 35$ pc. All simulations were run up to an age
of $T=13.5$ Gyr and final cluster models were calculated by taking 10 snapshots from the simulations centered around the age of each
globular cluster. The combined snapshots of the $N$-body clusters were scaled in mass and radius to match the density and velocity dispersion profiles of the
observed globular clusters and the best-fitting model was determined by interpolating in the grid of $N$-body simulations.

In the simulations done by \citet{baumgardt2017}, cluster stars followed \citet{kroupa2001} mass functions initially. However, observed present-day
mass functions of globular clusters show that many globular clusters have mass functions with significantly fewer low-mass stars than predicted by 
a Kroupa mass function
\citep[e.g.][]{demarchietal2007,paustetal2010,webbleigh2015,sollimabaumgardt2017}. This can be understood as a result of ongoing cluster dissolution which preferentially
removes low-mass stars from the clusters \citep{vesperiniheggie1997,baumgardtmakino2003}.
Since the clusters in the simulations of \citet{baumgardt2017} were isolated, they did not lose stars during their evolution and their mass
functions did not become depleted in low-mass stars. These models can therefore not be used to fit the mass functions of most Galactic globular clusters. 
In this paper we therefore 
ran a large set of additional simulations with initial mass functions (IMFs) depleted in low-mass stars to be able to match the observed mass functions of globular clusters
and derive more accurate estimates of the cluster parameters like the structural
parameters and total masses. The initial mass functions were set up as a combination of five
connected power-laws $N(m) \sim m^{\alpha}$ between mass limits of 0.1 and 15 M$_\odot$ and with mass function breaks at 
0.2 M$_\odot$, 0.5 M$_\odot$, 0.8 M$_\odot$ and 1.0 M$_\odot$. Table~\ref{tab:mfrange} gives an overview of the mass limits of the different power-law segments 
and the individual mass function slopes. The
slopes were derived by fitting power-laws MFs to the stellar mass functions of the $N$-body simulations from \citet{baumgardtsollima2017},
as well as the Monte Carlo simulations of \citet{askaretal2017} using the maximum-likelihood approach described
in \citet{clausetetal2009} and \citet{khalajbaumgardt2013} to calculate the best-fitting power-law exponents $\alpha$ for the different segments. \citet{baumgardtsollima2017} simulated star clusters in a circular orbit around
a central galaxy that was modeled as an isothermal sphere with a circular velocity of $v_c=220$ km/sec. Their simulations took the full tidal
field of the parent galaxy into account. The Monte Carlo simulations of \citet{askaretal2017}
assumed a point-mass galaxy with the same circular velocity. 
If fitted by a single power-law in the range $0.2 < m < 0.8$ $M_\odot$, the simulated 
models have global mass functions slopes of $\alpha=-1.5$ (corresponding to a Kroupa IMF, model 1), $\alpha=-1.0$ (model 2), $\alpha=-0.5$ (model 3),  $\alpha=0.0$ (model 4)
and $\alpha=+0.5$ (model 5).
Since our clusters are isolated, they undergo only very little mass loss during their evolution. Hence the initial mass functions are more or less equal to the final ones except at the high mass end 
where stellar evolution turns massive stars into compact remnants.


We linearly interpolated between the simulations, varying initial cluster concentration, initial half-mass radius and cluster mass function to find the cluster model that simultaneously 
provides the best fit to the surface density profile, velocity dispersion profile
and the mass functions at different radii of each individual globular cluster. In order to compare with the observed mass functions, we determined
the sky location and exact boundaries of the HST fields used to derive the stellar mass functions from the {\tt MAST} archive\footnote{\url{https://archive.stsci.edu/}},
projected the model clusters onto the sky and selected stars in the same area as the observed data.

For clusters which did not have measured mass functions, we estimated the global mass function based on the clusters' relaxation time (see discussion in sec.~4.2). 
We then performed only a 2D fit in our grid for these clusters, varying only the initial radius and cluster concentration but keeping the mass function fixed. Since the relaxation time
is determined from the fit itself, we did the fitting iteratively for clusters without a direct mass function measurement until a stable solution was obtained.

We applied the above procedure to all clusters except NGC 2419, where the radial velocity dispersion profile could not be
fitted with isotropic $N$-body models since it drops too quickly with radius in the outer parts. This
is most likely due to a radially anisotropic velocity dispersion profile \citep{ibataetal2011}, that could have been created during cluster
formation and was not erased by dynamical evolution due to the long relaxation time of NGC 2419. We therefore fitted NGC 2419
using radially anisotropic \citet{king1962} profiles. The profiles were created by the distribution function fitting method described 
in \citet{hilkeretal2007} using radially anisotropic Osipkov-Merritt models \citep{osipkov1979,merritt1985}. As Fig.~\ref{fig2a} shows,
the best-fitting cluster model found this way reproduces the observed velocity dispersion and surface density profile of NGC~2419 very well.
Our solution for $\omega$ Cen is based on the best-fitting no IMBH model from \citet{baumgardtetal2018}, who, in addition to the $N$-body simulations 
presented in this paper, have run an additional grid of simulations which vary the assumed retention fraction of stellar-mass black holes.
Since they lack radial velocity information, masses and structural parameters for NGC 6101 and NGC 6254 were determined by fitting the 
absolute number of main sequence stars at different radii which \citet{sollimabaumgardt2017} derived from the ACS Treasure project data.

\section{Results}

\subsection{47 Tuc and M15}

\begin{figure*}
\begin{center}
\includegraphics[width=0.95\textwidth]{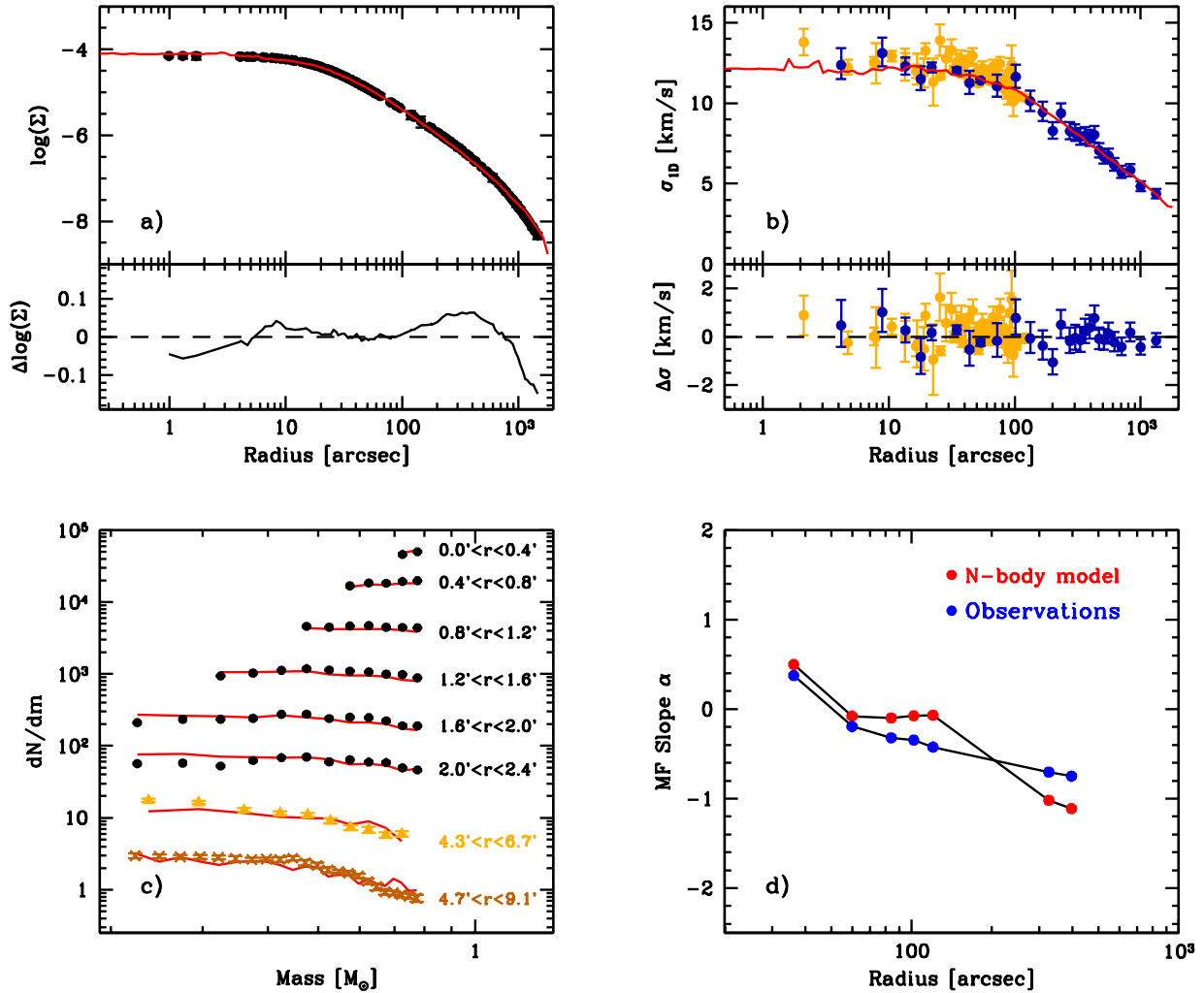}
\end{center}
\caption{Fit of the surface density profile (panel a), velocity dispersion profile (panel b), number of main-sequence stars as a function of stellar mass at 8 different
radii in the cluster (panel c) and mass function slope as a function of radius (panel d) of 47 Tuc. In each panel the best-fitting $N$-body model is shown by red lines or dots
while the observed data is shown in other colors. In panel b), the velocity dispersion profile based on proper motions is shown by orange circles while blue circles
show the radial velocity dispersion profile. The best-fitting $N$-body model is within 10\% of the observed surface density profile, within 0.5 km/sec of the observed
velocity dispersion profile and within $\Delta \alpha=0.3$ in mass function slope over the whole range of radii. In addition there is very good agreement in the absolute number of
main sequence stars at different radii between $N$-body model and the observed 47 Tuc.}
\label{47tucfit}
\end{figure*}

We start the discussion of our results by presenting the solution for the best-fitting models of 47 Tuc and M15, two of the best observed clusters in our sample, in greater detail. Fig. \ref{47tucfit} depicts the surface density profile (panel a), velocity dispersion
profile (panel b), mass distribution of main-sequence stars at different radii (panel c) and the slope of the best-fitting power-law to the stellar mass function as a 
function of radius (panel d) of 47 Tuc and our best-fitting $N$-body model.
In all panels, the $N$-body model is shown by red lines or circles while circles in other colors show the observed cluster data.
In panel b), blue circles show the observed radial velocity dispersion profile while orange circles show the proper motion dispersion profile from \citet{watkinsetal2015a}.
The number of stars as a function of stellar mass at different radii depicted in panel c) is taken from \citet{sollimabaumgardt2017} for radii inside $r=2.4'$, and \citet{demarchiparesce1995b} and \citet{richer2017}
for larger radii. It can be seen that the best-fitting $N$-body model reproduces the observed parameters of 47~Tuc very well. The surface density profile of the best-fitting
$N$-body model is within 10\% of the observed surface density profile for all radii except near the tidal radius. Similarly the velocity dispersion profiles of
the $N$-body model and 47 Tuc agree to within 0.5 km/sec at all radii between 1'' $< r <$ 1000''. The reduced $\chi_r^2$ value given as the sum of the error weighted velocity differences
normalized by the number of data points is almost exactly 1, indicating excellent agreement (see Table~2). The absolute number of stars as a function of 
mass at different radii in panel c) is also in very good agreement between $N$-body model and observed cluster. Since the velocity 
dispersion profile essentially determines the total cluster mass, a good
agreement in the absolute number of main sequence stars means that the $N$-body model must also have the same amount of mass in compact remnants as 
the real 47 Tuc, i.e. our chosen mass function must be a good description of the mass function of 47 Tuc.
The best-fitting $N$-body model also has about the same amount of mass segregation as 47 Tuc, since at all radii the observed mass function slope is within  $\Delta \alpha=0.3$
of the mass function slope of the best-fitting $N$-body model. 

\begin{figure*}
\begin{center}
\includegraphics[width=0.95\textwidth]{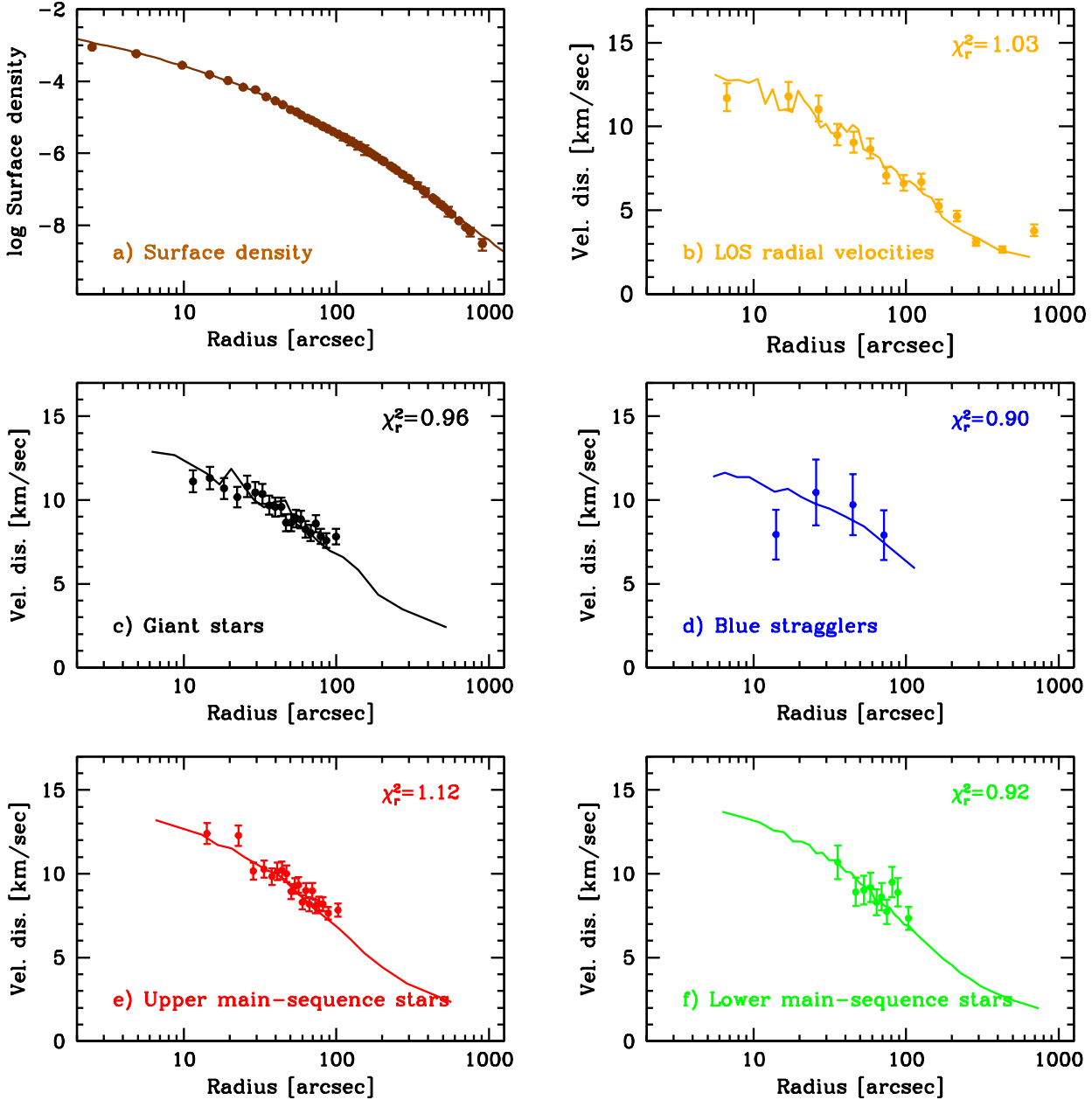}
\end{center}
\caption{Fit of the surface density profile (panel a), radial velocity dispersion profile (b), and the proper motion dispersion profiles of giant stars (c), blue stragglers
(d), upper main sequence stars (e) and lower main sequence stars (f) of M~15. Observed data is shown as solid points with error bars while the prediction of the best-fitting
$N$-body model is shown as solid lines in each panel. The proper motion dispersion profiles were calculated from the HST data published by \citet{bellinietal2014}.
The reduced $\chi_r^2$ values from the comparison of the theoretical and observed profiles are given in the upper right corner of each panel. All $\chi_r^2$ values are
close to unity, indicating excellent agreement between the modeled cluster and the observations.}
\label{m15fit}
\end{figure*}

Fig. \ref{m15fit} shows a comparison of the surface and velocity dispersion profile of M15 with the prediction of our best-fitting $N$-body model. Shown are the
surface density profile (top left), the radial velocity dispersion profile (top right panel), and the proper motion dispersion profiles of giant
stars, blue stragglers, upper main sequence stars and lower main sequence stars. The proper motion dispersion profiles were calculated using
the data published by \citet{bellinietal2014}, restricting ourselves to stars outside the central 10'' with proper motion errors less 
than 3 km/sec and absolute velocities within 35 km/sec of the cluster mean. 
We define upper main sequence stars as stars with magnitudes $19 < F_{814W} < 20$ and $F_{814W}-F_{660W}>0.4$ and lower main sequence 
stars as stars with $20 < F_{814W}$ and $F_{814W}-F_{660W}>0.4$ in the catalogue of \citet{bellinietal2014}. These limits translate roughly into mass limits of
$0.75 > m > 0.70$ M$_\odot$ and $m<0.70$ M$_\odot$ for upper and lower main sequence stars respectively. For the comparison with the blue stragglers,
which are not present in our $N$-body models, we use massive white dwarfs with masses $1.0 < m < 1.4$ M$_\odot$.

The observations of \citet{sollimabaumgardt2017}
show that M15 is strongly mass segregated since the stellar mass function changes from $\alpha=0.2$ in the centre to $\alpha=-1.2$
outside the clusters' half-mass radius. Furthermore, the surface density of M15 is strongly increasing towards the centre
down to the smallest radius for which it can be measured, indicating that M15 is in or past core collapse. The mass segregation 
of M15 is also evident in the dependency of the velocity dispersion profile with stellar mass, the observed 
velocity dispersion at a projected radius of $r=30''$ for example changes from $\sigma \approx 10$ km/sec for blue stragglers to $\sigma = 12$ km/sec for lower
main sequence stars. Although small, this change is clearly resolved in the observations. Our best-fitting 
$N$-body model is again in excellent agreement with the observed data on M15 since it fits the surface density profile of the cluster and
the proper motion dispersion profile of each stellar mass group. In addition the $N$-body model reproduces the line-of-sight velocity dispersion profile, since
the reduced $\chi^2$ values of the fits to the various velocity dispersion profiles all being very close to unity.

\subsection{Cluster masses and structural parameters}

\begin{figure*}
\begin{center}
\includegraphics[width=0.9\textwidth]{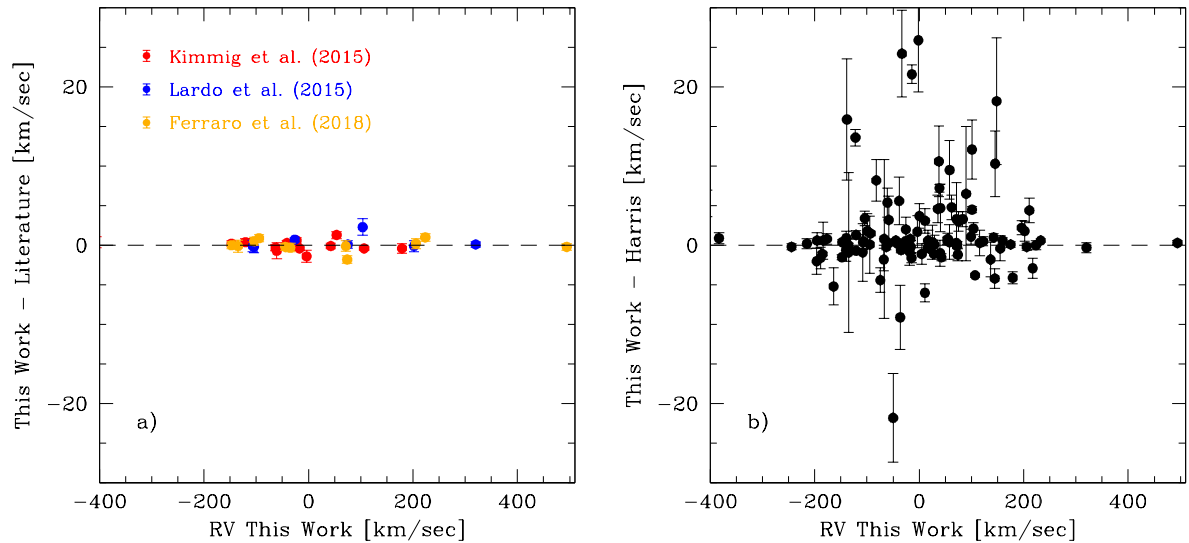}
\end{center}
\caption{Comparison of the mean cluster velocities derived here with the radial velocities determined by \citet{kimmigetal2015}, \citet{lardoetal2015} and \citet{ferraroetal2018} 
(left panel) and the mean cluster velocities given in the Harris catalogue (right panel). Our data generally agrees to within 1 km/sec with recent literature values.
Clusters which show larger differences in the Harris catalogue are mostly clusters where the radial velocities in the Harris catalogue are based on low-resolution spectroscopy
and have relatively large error bars.}
\label{meanvr}
\end{figure*}

Table~2 presents the masses and structural parameters of all Galactic globular clusters for which we could derive radial velocity dispersion profiles. 
For each cluster we determined the number of member stars, the mean radial velocity and its error, the total cluster mass and $M/L$ ratio, the core radius, 3D half-mass radius and projected half-light radius,
the average density within the core and half-mass radius, the half-mass relaxation time, the global mass function slope between mass limits of 0.2 to 0.8 M$_\odot$, the one-dimensional, mass-weighted, central velocity dispersion, and the central escape velocity.
The mass function slopes were either calculated from the $N$-body model that best reproduces the observed  mass functions (for cluster that have measured mass functions) or using the relation between 
relaxation time and mass function slope that will be discussed in sec. 4.3.
We count as radial velocity members all stars with membership probability $P_i>0.01$ in each cluster. For some bulge clusters with strong field star
contamination like Ter~5 these numbers might overestimate the number of true member stars.
The mean radial velocities were calculated as the weighted mean of the individual radial velocities of all cluster members, except for
NGC~6101, NGC~6293 and NGC~6584 where we took the mean radial velocity from \citet{kamannetal2018} (NGC 6293) or \citet{harris1996} (NGC~6101 and NGC~6584).
Fig.~\ref{meanvr} compares the mean radial velocities which we have determined with the radial velocities 
determined by \citet{kimmigetal2015}, \citet{lardoetal2015} and \citet{ferraroetal2018} from resolved spectroscopy of individual member stars (left panel)
and the radial velocities given by Harris (1996) (right panel). It can be
seen that the mean radial velocities derived here generally agree to within 1~km/sec with the mean radial velocities determined from individual stars. The remaining differences
are probably due to uncertainties in the absolute wavelength calibration of the individual spectra, which are difficult to quantify but, judging from Fig.~\ref{meanvr}, could be around 1~km/sec.
For most clusters our radial velocities are also in good agreement with the values given by Harris (1996). Clusters where the radial velocities differ more strongly
are usually not very well studied clusters where the mean radial velocities have relative large error bars in the Harris catalogue.

Clusters that have a zero $\chi^2_r$ value in column 4 of Table~\ref{tabresult} are clusters with few member stars for which we grouped the stars in only one or two velocity bins, which 
are then reproduced almost exactly by the best-fitting $N$-body model. Column 5 of Table~\ref{tabresult} lists the cluster distances. For clusters for which \citet{watkinsetal2015a} have determined proper motion dispersion profiles,
we vary the cluster distance until we obtain the best fit to the combined proper motion and radial velocity dispersion profile. For all other clusters the cluster
distance is taken from the literature. 
Interestingly our best-fitting distance for 47 Tuc ($d=4.41$~kpc) is now in much better agreement
with other distance methods that generally find a cluster distance around $d=4.5$~kpc \citep[e.g.][]{grattonetal2003,bonoetal2008,dotteretal2010,woodleyetal2012} 
than the kinematic distances derived by \citet{baumgardt2017} ($d=3.95 \pm 0.05$~kpc) or \citet{watkinsetal2015b} ($d=4.15 \pm 0.08$~kpc). 
The reason is that the HST proper motion dispersion profile of \cite{watkinsetal2015a} measures velocities only in the
central 100'' while radial velocities are available mainly for stars further away from the centre. This together with the fact that
the new best-fitting model for 47 Tuc has fewer low-mass stars than the Kroupa mass function used by \citet{baumgardt2017}, which
lowers the velocity dispersion in the outer parts compared to the centre, pushes the best-fitting cluster distance to a larger value
and brings it into much better agreement with the other methods.
This again shows the importance of correctly modeling mass segregation when fitting models to observed clusters.

We calculated the core radius and average density inside the core by applying eq. 149 in \citet{spitzer1987} iteratively to the $N$-body data until a
stable solution was found, using a correction factor of 0.517 in the conversion of core density to central density as described in
\citet{baumgardtetal2003b}. From all stars inside the core radius we then calculated the three-dimensional central velocity dispersion,
weighting the individual stellar velocities with the masses of the stars. From the fastest single stars inside the core we also calculated the
escape velocities of the clusters given in the final column of Table~2.
Figs. \ref{fig1a} to \ref{fig15a} depict our fits to the observed surface density and velocity dispersion profiles for clusters
with more than 100 member stars. It can be seen that we generally obtain very good fits to both profiles. The differences in
the surface density profiles over most parts of the clusters are usually within 15\%. Only in the very centre or near the tidal radius
one can sometimes see larger differences. Observational uncertainties might be a reason for the differences since the observed surface
density profiles could be affected by low-N noise in the core and uncertainties about the density of background stars near the tidal radius. The
velocity dispersion profiles usually also agree to within 1 km/sec. Remaining differences could be due to a variety of reasons like stellar
binaries, orbital anisotropy and tidal effects. Stellar binaries in particular could inflate the velocity dispersion profiles of
low-mass clusters if present in sufficient numbers \citep[e.g.][]{blechaetal2004,gielesetal2010}. In addition tidal effects \cite[e.g.][]{kuepperetal2010} 
or non-members could be responsible for the higher than predicted velocity dispersion profiles seen in the outer parts of some clusters like NGC~1851.
Using proper motions from the {\tt GAIA} satellite will help removing non-members from our data and assess whether the deviations seen
are real or due to interlopers.
\begin{figure*}
\begin{center}
\includegraphics[width=\textwidth]{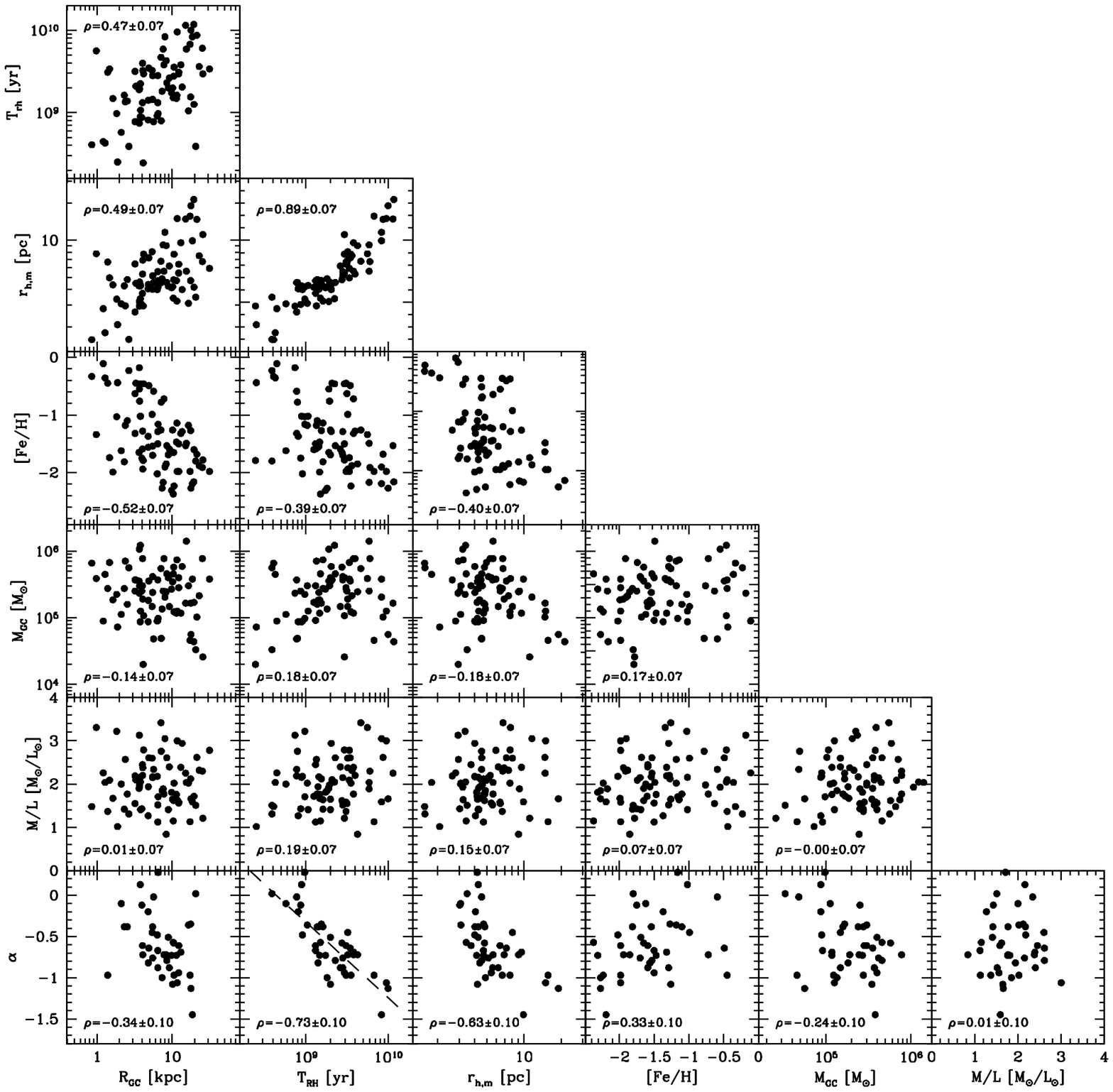}
\end{center}
\caption{Univariate correlations between the cluster mass functions $\alpha$, galactocentric distances $R_{GC}$, half-mass relaxation times $T_{RH}$, half-mass radii $r_{h,m}$, cluster masses $M_{GC}$
and mass-to-light ratios $M/L$ for all studied clusters. In each panel we also show the Spearman rank order coefficient $\rho$ and its $1\sigma$ error. The dashed line in the $\alpha$ vs. $T_{RH}$
plane shows the best-fitting linear relation to the cluster distribution.}
\label{allesgegenalles}
\end{figure*}

\subsection{Correlations between cluster parameters}

We next discuss several univariate correlations that we found among the cluster parameters. Fig.\ \ref{allesgegenalles} depicts correlations between the cluster mass functions 
$\alpha$, galactocentric distances $R_{GC}$, half-mass relaxation times $T_{RH}$, half-mass radii $r_{h,m}$, cluster metallicities [Fe/H], cluster masses $M_{GC}$,
and the mass-to-light ratios $M/L$ for all clusters with less than 30\% relative mass error.
Mass functions are only available for 47 clusters, i.e. about one third of all galactic globular clusters, which could introduce some bias into the cluster distribution since
the selection of clusters that have measured mass functions is not random but weighted towards nearby and massive clusters. For all other cluster parameters 
our sample includes more than 2/3 of all known galactic globular clusters, making a bias 
much less likely, especially for more massive globular clusters with $M_V<-8.0$
where our sample is almost 100\% complete. Each panel also shows the Spearman rank order coefficient $\rho$ and its $1\sigma$ error. Our data shows a clear correlation between the
galactocentric distance of a globular cluster and its half-mass radius. Part of this correlation could be due to the stronger tidal field in the inner parts of the Milky Way,
which limits the size of globular clusters to smaller values. \citet{vandenberghetal1991} found a similar relation between
galactocentric distance and the projected half-light radius. We confirm their results for the half-mass radius. 
\begin{figure*}
\begin{center}
\includegraphics[width=\textwidth]{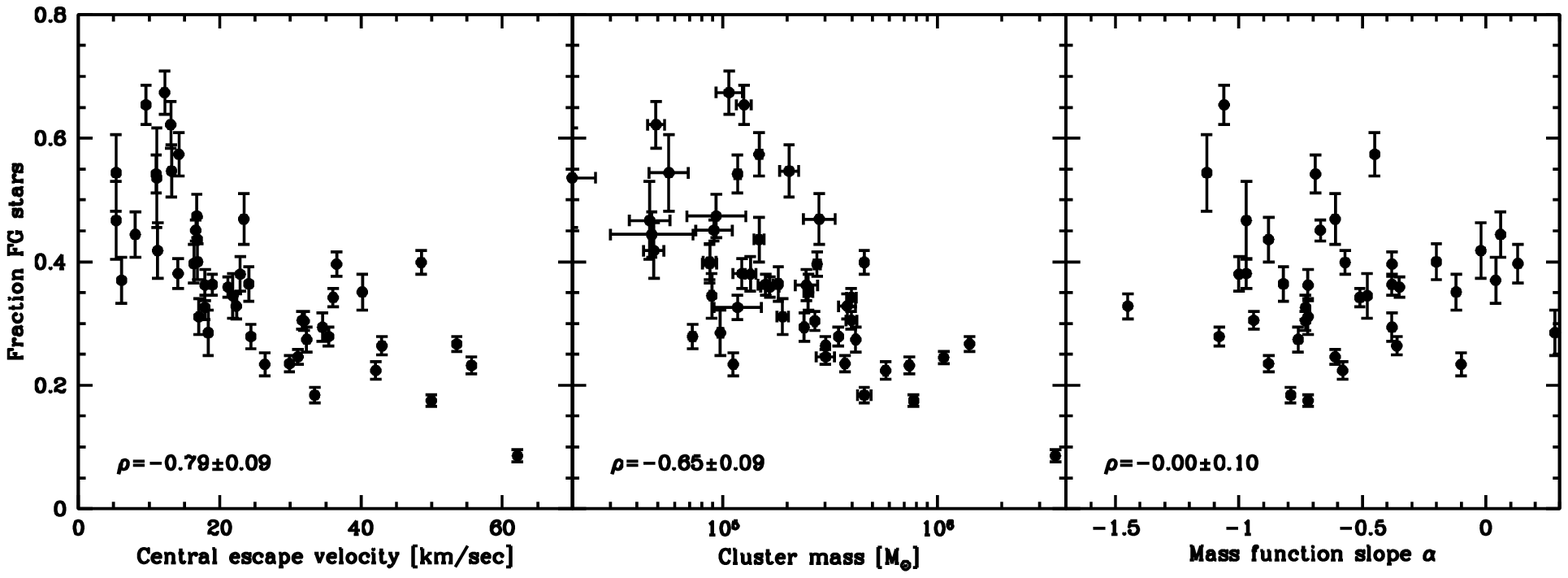}
\end{center}
\caption{Correlation of the fraction of first generation stars as determined by \citet{miloneetal2017} with the escape velocity (left panel), the cluster mass (middle) and the mass function slope (right
panel). The fraction of first generation stars in a globular cluster shows the strongest dependency on the escape velocity from the center of a globular cluster and shows no dependency
on the mass function slope. This could indicate that the ability of a globular cluster to retain stellar ejecta is the most important parameter determining the fraction of second generation stars
and not its later mass loss.}
\label{milone}
\end{figure*}

\citet{sollimabaumgardt2017} found a correlation between the
mass function slope of 35 globular clusters from the {\it Globular Cluster ACS Treasury Project} and their half-mass relaxation times in the sense that clusters 
with more positive mass function slopes (fewer low-mass stars) have smaller relaxation times. We confirm this correlation with our larger cluster sample.
The reason for the existence of this correlation could be that clusters in stronger tidal fields experience more mass loss while at the same time the stronger tidal field 
limits them to a smaller volume and therefore radius. In addition clusters with shorter relaxation times develop mass segregation in a shorter amount of time, leading to
a stronger depletion of low-mass stars by the tidal field of the Milky Way.
Dynamical evolution could also explain the correlation of mass function slope with galactocentric distance, although this correlation is weaker then the correlation
with the relaxation time. The mass function also correlates with both the half-mass 
radius and, less strongly, with the mass of a globular cluster. 

Most correlations between the cluster metallicity and other cluster parameters can be explained by the fact that high metallicity clusters are related to the Galactic disc and bulge and are
therefore on average closer to the Galactic centre than halo clusters \citep{zinn1985}.
They therefore experience a stronger tidal field which makes them more compact and also reduces their relaxation times. The strong correlation between metallicity
and mass function slope could therefore be a result of dynamical evolution and might not reflect an initial variation, although we cannot rule out such a variation either.
The $M/L$ ratios do not show a significant correlation with any of the other cluster parameters. In particular we 
do not find a correlation between $M/L$ ratio and either metallicity or mass function slope. Such correlations should in principle exist since for example the loss of 
low-mass stars decreases the $M/L$ ratio of a globular cluster \citep{baumgardtmakino2003,kruijssenmieske2009}. In addition, the loss of stellar remnants can
decrease the $M/L$ ratio of a globular cluster \citep{bianchinietal2017}. The reason for the absence
of a correlation could be that the resulting change in $M/L$ ratio is too small to cause a noticeable difference or is compensated for by the correlation between mass
function slope and cluster metallicity.

The dashed line in the $\alpha$ vs. $T_{RH}$ plot shows the best-fitting linear relation to the observed globular cluster distribution which is given by $\alpha = 8.23 \pm 1.10 - (0.95 \pm 0.11) \; \log T_{RH}$.
We use this relation to infer the mass function of globular clusters for which no direct measurement is available (see discussion in sec. 3).

Fig. \ref{milone} finally depicts the fraction of first generation (FG) stars in globular clusters found by \citet{miloneetal2017} through high precision HST photometry of red giant branch stars
as a function of the central escape velocity of a cluster (left panel), the cluster mass (middle panel) and the global mass function slope between 0.2 and 0.8 M$_\odot$
(right panel). It can be seen that the fraction of
first generation stars anti-correlates strongly with either the cluster mass (confirming the results found by Milone et al. 2017) or the central escape velocity. 
In particular clusters with small central escape velocities consist predominantly
of first generation stars while the fraction of these stars drops to about 20\% for clusters with escape velocities $v_{esc}>40$ km/sec. However, as the right panel of Fig. \ref{milone} shows,
the fraction of FG stars shows no correlation with the global mass function slope of a globular cluster. If clusters formed with a low fraction of second generation (SG) stars which later increases
due to the loss of less centrally concentrated first generation stars \citep[e.g.][]{dantonacaloi2008}, one would however expect that such a relation should be established since two-body relaxation will push
low-mass stars to the outer cluster parts where they are preferentially lost. Hence clusters that have lost a large fraction of first generation stars should also have lost a higher
fraction of their low-mass stars and should therefore have more evolved mass functions. As Fig. \ref{milone} shows such a correlation is not observed, calling into question the strong mass-loss 
scenario. If mass loss of first generation stars has increased the fraction of second generation stars, then this mass loss must have happened very early in the cluster evolution before
dynamical mass loss has set in, possible through e.g. gas expulsion \citep{decressinetal2010,khalajbaumgardt2015} or giant molecular cloud encounters
\citep{kruijssenetal2012}. Still such a scenario would not be able to explain the strong dependency of the fraction of FG stars on the central escape
velocity, which could instead indicate that the formation efficiency of SG stars depends on the ability of a cluster to retain the stellar wind ejecta from polluting stars: For low escape velocities
of $v_{esc}=10$ km/sec or less nearly all ejecta leave the cluster and only few SG stars are formed, while at escape velocities larger than $v_{esc}=40$ km/sec all ejecta are kept,
meaning that the fraction of SG stars becomes constant. \citet{georgievetal2009} also suggested that
the central escape velocity can be used to describe the degree of self
enrichment in a globular cluster based on the correlation of escape
velocity with metallicity for globular clusters containing extended
horizontal branches.

\section{Conclusions}

We have determined individual stellar radial velocities of more than $15,000$ stars in 90 globular clusters from archival ESO/VLT and Keck spectra. Combining this data with published literature velocities we 
then calculated radial velocity dispersion profiles of globular clusters using a maximum-likelihood approach. A comparison of these velocity dispersion profiles together with the 
surface density profiles of the globular clusters and the stellar mass functions recently determined by \citet{sollimabaumgardt2017} with a large set of $N$-body models then allowed us to
determine total cluster masses, global mass function slopes and the structural parameters of 112 Galactic globular clusters. Our new cluster sample is more than twice as large as the 
sample studied
by \citet{baumgardt2017} and includes over 2/3 of all globular clusters in the Milky Way. It is essentially complete for all globular clusters more massive than $2 \cdot 10^5$ M$_\odot$. 
In addition we are now able to fit the stellar mass functions at different radii in a large sample of globular clusters. Including the stellar 
mass functions in the fit significantly increases the accuracy of the derived cluster parameters
since it allows us to better model the internal mass distribution of a globular cluster. $N$-body simulations are therefore another way to model
globular clusters in addition to fitting them using multi-mass King Michie models 
\citep[e.g.][]{sollimaetal2012}, {\tt LIMEPY} models \citep{gielesetal2018} or Monte Carlo simulations \citep[e.g.][]{gierszheggie2009,gierszheggie2011}.

Our data shows several interesting correlations among the globular cluster parameters. We confirm an increase of the cluster sizes with galactocentric distance first found by \citet{vandenberghetal1991}
as well as the correlation between mass function slope and cluster relaxation time found by \citet{sollimabaumgardt2017}. The latter is possibly a result of mass segregation and dynamical mass loss
of globular clusters that are evolving in tidal fields of different strengths. We also find a strong correlation between the central escape velocity from a globular cluster and the fraction of second generation stars
found by \citet{miloneetal2017}  but no correlation between this fraction and the mass function slope of the clusters. These could indicate that the fraction of second generation stars is determined
by conditions prevalent at the formation of globular clusters and not by their later dynamical evolution.
For any self enrichment scenario to work, one would
however need rather strongly depleted initial mass functions, since
canonical Kroupa or Chabrier mass functions produce an SG star
fraction significantly lower than observed
 
\section*{Acknowledgments}

We thank Andreas Koch, Nikolay Kacharov and Sabine M\"ohler for sharing unpublished radial velocities of stars in NGC~288 and M75 with us and Harvey Richer for sending us
unpublished mass function data. We thank Mirek Giersz for useful discussions and his help with the MOCCA simulations which helped improve our fitting method. We 
also acknowledge the help of Sarah Sweet with the reduction of the {\tt DEIMOS} data and Tony Sohn and Emily Cunningham 
for providing us with telluric template spectra that were useful in deriving radial velocities from the {\tt DEIMOS} data. 
We finally thank Ivo Saviane for sharing his {\tt FORS2} spectra of stars in several clusters with us.
This paper is based on data obtained from the ESO Science Archive Facility. This research has also
made use of the Keck Observatory Archive (KOA), which is operated by the W. M. Keck Observatory and the NASA Exoplanet Science Institute (NExScI), under contract with the National 
Aeronautics and Space Administration. Some of the data presented in this paper was obtained from the Mikulski Archive for Space Telescopes (MAST). STScI is operated by the Association 
of Universities for Research in Astronomy, Inc., under NASA contract NAS5-26555. Support for MAST for non-HST data is provided by the NASA Office of Space Science via grant NNX09AF08G 
and by other grants and contracts. This research has made extensive use of the SIMBAD database,
operated at CDS, Strasbourg, France.

\bibliographystyle{mn2e}
\bibliography{mybib}

\begin{landscape}
\begin{table}
\caption{Derived parameters of all studied globular clusters. For each cluster, the Table gives the name of the cluster, the number of cluster members with measured radial velocities, the mean radial velocity of the cluster and its error,  the reduced $\chi_r^2$ value of the velocity dispersion fit, the cluster distance (either from literature data or obtained as part of the fitting process), the total cluster mass and V-band $M/L$ ratio, the core, half-mass and projected half-light radius, the central density and the density inside the half-mass radius,the half-mass relaxation time, the global mass function slope, the mass-weighted central (1D) velocity dispersion, and the central escape velocity.}

\begin{tabular}{lr@{\hspace*{0.2cm}}rcrrcrrr@{}c@{\hspace*{-0.15cm}}c@{}rrrr}
\hline
 & & & & & & & & & & & & & & \\[-0.3cm]
\multirow{2}{*}{Name} & \multirow{2}{*}{$N_{RV}$} & \multicolumn{1}{c}{$<\! v_r \! >\!\!\!\!$} & \multirow{2}{*}{$\chi^2_r$} &\multicolumn{1}{c}{Dist.} & \multicolumn{1}{c}{Mass} & \multicolumn{1}{c}{$M/L_V$} & \multicolumn{1}{c}{$r_c$} & \multicolumn{1}{c}{$r_{h,m}$} & \multicolumn{1}{c}{$r_{h,lp}$} & \multicolumn{1}{c}{$\log \rho_{c}$} & \multicolumn{1}{c}{$\log \rho_{hm}$} & \multicolumn{1}{r}{$\log T_{RH}\!\!\!\!$} & \multicolumn{1}{c}{MF$\!$} & \multicolumn{1}{c}{$\sigma_0\!\!\!\!\!$} & \multicolumn{1}{c}{$v_{esc}\!\!\!\!\!$}\\
 & & \multicolumn{1}{c}{[km/sec]$\!\!\!\!$} & & \multicolumn{1}{c}{[kpc]} & \multicolumn{1}{c}{[M$_\odot$]} & \multicolumn{1}{c}{[M$_\odot$/L$_\odot$]}  & \multicolumn{1}{c}{[pc]} & \multicolumn{1}{c}{[pc]} & \multicolumn{1}{c}{[pc]} & \multicolumn{1}{c}{[M$_\odot/pc^3$]} &  \multicolumn{1}{c}{[M$_\odot/pc^3$]} & \multicolumn{1}{c}{[yr]$\!\!\!\!$} & \multicolumn{1}{c}{Slope$\!$} & \multicolumn{1}{c}{[km/sec]$\!\!\!\!\!$} & \multicolumn{1}{c}{[km/sec]$\!\!\!\!\!$}\\[+0.1cm]
\hline
 & & & & & & & & & & & &  & & \\[-0.3cm]
NGC 104 & 2896 &  -17.2 $\pm$  0.1 & 0.98 & 4.41$^b$ & $7.79 \pm 0.05 \cdot 10^5$ &  $1.77 \pm 0.16$ & 0.49 & 5.62 & 3.57 & 5.31 &  2.72 & 9.58 & -0.53$^c$ & 12.2 & 49.9\\
NGC 288 & 584 &  -44.8 $\pm$  0.1 & 1.34 & 9.80$^b$ & $1.16 \pm 0.03 \cdot 10^5$ &  $2.39 \pm 0.17$ & 4.77 & 9.54 & 6.99 & 1.62 &  1.22 & 9.58 & -0.45$^c$ & 3.3 & 11.0\\
NGC 362 & 487 &  223.5 $\pm$  0.3 & 0.98 & 9.40$^b$ & $3.45 \pm 0.05 \cdot 10^5$ &  $1.65 \pm 0.25$ & 0.47 & 4.21 & 2.32 & 4.50 &  2.75 & 9.30 & -0.89$^c$ & 8.8 & 35.4\\
NGC 1261 & 231 &   71.5 $\pm$  0.3 & 2.17 & 15.49$^a$ & $1.67 \pm 0.17 \cdot 10^5$ &  $2.12 \pm 0.38$ & 1.19 & 4.72 & 3.08 & 3.33 &  2.33 & 9.19 & -0.53$^c$ & 5.6 & 21.2\\
\hline
\end{tabular}
\end{table}

\end{landscape}

\begin{landscape}
\begin{table}
\contcaption{}
\begin{tabular}{l@{\hspace*{-0.05cm}}rrcrrcrrr@{}c@{\hspace*{-0.15cm}}c@{}rrrr}
\hline
 & & & & & & & & & & & & & & \\[-0.3cm]
\multirow{2}{*}{Name} & \multirow{2}{*}{$N_{RV}$} & \multicolumn{1}{c}{$<\! v_r \! >\!\!\!\!$} & \multirow{2}{*}{$\chi^2_r$} &\multicolumn{1}{c}{Dist.} & \multicolumn{1}{c}{Mass} & \multicolumn{1}{c}{$M/L_V$} & \multicolumn{1}{c}{$r_c$} & \multicolumn{1}{c}{$r_{h,m}$} & \multicolumn{1}{c}{$r_{h,lp}$} & \multicolumn{1}{c}{$\log \rho_{c}$} & \multicolumn{1}{c}{$\log \rho_{hm}$} & \multicolumn{1}{r}{$\log T_{RH}\!\!\!\!$} & \multicolumn{1}{c}{MF$\!$} & \multicolumn{1}{c}{$\sigma_0\!\!\!\!\!$} & \multicolumn{1}{c}{$v_{esc}\!\!\!\!\!$}\\
 & & \multicolumn{1}{c}{[km/sec]$\!\!\!\!$} & & \multicolumn{1}{c}{[kpc]} & \multicolumn{1}{c}{[M$_\odot$]} & \multicolumn{1}{c}{[M$_\odot$/L$_\odot$]}  & \multicolumn{1}{c}{[pc]} & \multicolumn{1}{c}{[pc]} & \multicolumn{1}{c}{[pc]} & \multicolumn{1}{c}{[M$_\odot/pc^3$]} &  \multicolumn{1}{c}{[M$_\odot/pc^3$]} & \multicolumn{1}{c}{[yr]$\!\!\!\!$} & \multicolumn{1}{c}{Slope$\!$} & \multicolumn{1}{c}{[km/sec]$\!\!\!\!\!$} & \multicolumn{1}{c}{[km/sec]$\!\!\!\!\!$}\\[+0.1cm]
\hline
 & & & & & & & & & & & &  & & \\[-0.3cm]
ESO 452-SC11 & 13 &   16.4 $\pm$  0.4 & 0.00 & 7.80$^a$ & $1.06 \pm 0.57 \cdot 10^4$ &  $3.38 \pm 1.91$ & 0.16 & 5.21 & 3.66 & 4.06 &  1.02 & 8.73 & -0.05$^d$ & 1.6 & 6.3\\
NGC 6205 & 386 & -244.4 $\pm$  0.3 & 2.92 & 6.60$^a$ & $4.53 \pm 0.34 \cdot 10^5$ &  $2.61 \pm 0.26$ & 1.58 & 4.56 & 3.08 & 3.58 &  2.87 & 9.36 & -0.62$^c$ & 9.2 & 33.4\\
NGC 6229 & 21 & -138.3 $\pm$  0.8 & 0.00 & 30.62$^a$ & $2.91 \pm 0.94 \cdot 10^5$ &  $2.42 \pm 0.88$ & 1.16 & 5.01 & 3.18 & 3.56 &  2.46 & 9.35 & -0.63$^d$ & 7.1 & 27.3\\
NGC 6218 & 469 &  -41.2 $\pm$  0.2 & 0.99 & 5.22$^a$ & $8.65 \pm 0.60 \cdot 10^4$ &  $1.27 \pm 0.30$ & 0.71 & 4.05 & 2.82 & 4.05 &  2.51 & 8.91 & -0.25$^c$ & 4.5 & 16.9\\
\hline
\end{tabular}
\end{table}

\end{landscape}

\begin{landscape}
\begin{table}
\contcaption{}
\begin{tabular}{lr@{\hspace*{0.2cm}}rcrrcrrr@{}c@{\hspace*{-0.15cm}}c@{}rrrr}
\hline
 & & & & & & & & & & & & & & \\[-0.3cm]
\multirow{2}{*}{Name} & \multirow{2}{*}{$N_{RV}$} & \multicolumn{1}{c}{$<\! v_r \! >\!\!\!\!$} & \multirow{2}{*}{$\chi^2_r$} &\multicolumn{1}{c}{Dist.} & \multicolumn{1}{c}{Mass} & \multicolumn{1}{c}{$M/L_V$} & \multicolumn{1}{c}{$r_c$} & \multicolumn{1}{c}{$r_{h,m}$} & \multicolumn{1}{c}{$r_{h,lp}$} & \multicolumn{1}{c}{$\log \rho_{c}$} & \multicolumn{1}{c}{$\log \rho_{hm}$} & \multicolumn{1}{r}{$\log T_{RH}\!\!\!\!$} & \multicolumn{1}{c}{MF$\!$} & \multicolumn{1}{c}{$\sigma_0\!\!\!\!\!$} & \multicolumn{1}{c}{$v_{esc}\!\!\!\!\!$}\\
 & & \multicolumn{1}{c}{[km/sec]$\!\!\!\!$} & & \multicolumn{1}{c}{[kpc]} & \multicolumn{1}{c}{[M$_\odot$]} & \multicolumn{1}{c}{[M$_\odot$/L$_\odot$]}  & \multicolumn{1}{c}{[pc]} & \multicolumn{1}{c}{[pc]} & \multicolumn{1}{c}{[pc]} & \multicolumn{1}{c}{[M$_\odot/pc^3$]} &  \multicolumn{1}{c}{[M$_\odot/pc^3$]} & \multicolumn{1}{c}{[yr]$\!\!\!\!$} & \multicolumn{1}{c}{Slope$\!$} & \multicolumn{1}{c}{[km/sec]$\!\!\!\!\!$} & \multicolumn{1}{c}{[km/sec]$\!\!\!\!\!$}\\[+0.1cm]
\hline
 & & & & & & & & & & & &  & & \\[-0.3cm]
NGC 6626 & 201 &   11.0 $\pm$  0.6 & 2.02 & 5.50$^a$ & $3.69 \pm 0.38 \cdot 10^5$ &  $2.19 \pm 0.35$ & 0.13 & 2.63 & 1.76 & 7.42 &  3.67 & 8.89 & -0.18$^d$ & 12.6 & 49.5\\
NGC 6642 & 13 &  -33.0 $\pm$  1.0 & 0.00 & 8.05$^a$ & $8.72 \pm 5.49 \cdot 10^4$ &  $1.77 \pm 1.13$ & 0.12 & 1.99 & 1.50 & 6.34 &  3.89 & 8.36 & 0.30$^d$ & 6.9 & 27.5\\
NGC 6656 & 720 & -147.8 $\pm$  0.3 & 1.04 & 3.10$^b$ & $4.16 \pm 0.05 \cdot 10^5$ &  $2.15 \pm 0.08$ & 1.19 & 5.16 & 3.26 & 3.66 &  2.56 & 9.45 & -0.68$^c$ & 8.4 & 32.3\\
NGC 6681 & 32 &  217.4 $\pm$  0.9 & 1.12 & 9.20$^b$ & $1.13 \pm 0.02 \cdot 10^5$ &  $2.00 \pm 0.28$ & 0.05 & 3.07 & 2.10 & 7.33 &  3.07 & 8.76 & -0.00$^d$ & 7.1 & 26.4\\
\hline
\end{tabular}
\label{tabresult}
\begin{tabbing}
\hspace*{0.5cm} Notes: \= a: Cluster distance from literature, b: Fitted cluster distance, c: Global mass function slope determined from observed mass functions,\\
 \> d: Global mass function slope estimated from clusters' relaxation time\\
\end{tabbing}
\end{table}

\end{landscape}

\clearpage

\appendix

\section{ESO/VLT and Keck program IDs used to derive radial velocities of individual stars in globular clusters}

\begin{table}
\caption{ESO/VLT and Keck program IDs used to derive radial velocities of stars in globular clusters.}
\begin{tabularx}{0.9\columnwidth}{lll}
\hline
Cluster & Program IDs\\
\hline
Arp 2 & 075.D-0075(A), 075.D-0075(B)\\
E 3 & 097.D-0056(A)\\
HP 1 & 093.D-0124(A)\\
IC 4499 & 097.D-0111(A)\\
Lil 1 & 085.D-0377(A), 087.D-0716(A), 089.D-0306(A)\\
NGC 104 &  71.B-0414(A), 072.D-0777(A), 073.D-0760(A),\\
 & 074.B-0415(A), 075.D-0043(A), 081.D-0287(A),\\
\hline
\end{tabularx}
\label{tabspecsrc}
\end{table}

\begin{table}
\contcaption{}
\begin{tabularx}{0.9\columnwidth}{lll}
\hline
Cluster & Program IDs\\
\hline
NGC 5986 & 083.D-0530(A), 193.D-0232(F)\\
NGC 6093 &  65.L-0518(A), 073.D-0695(A), 079.D-0021(A),\\
 & 083.D-0530(A)\\
NGC 6121 & 073.D-0093(A), 081.D-0356(A), 083.B-0083(A),\\
 & 085.D-0205(A), 085.D-0537(A), 089.D-0062(A),\\
 & 089.D-0298(A), 093.D-0789(A), 093.D-0818(A),\\
 & 095.D-0819(A), 193.D-0232(B)\\
\hline
\end{tabularx}
\end{table}

\begin{table}
\contcaption{}
\begin{tabularx}{0.9\columnwidth}{lll}
\hline
Cluster & Program IDs\\
\hline
NGC 6752 & 073.D-0100(A), 075.D-0492(A), 079.D-0674(A),\\
 & 079.D-0674(B), 079.D-0674(C), 081.D-0356(A),\\
 & 083.B-0083(A), 089.D-0298(A), 091.D-0329(A),\\
 & 095.D-0320(A), 193.D-0232(B)\\
NGC 6809 & 073.D-0211(A), 091.D-0329(A), 093.D-0270(A),\\
 & 093.D-0818(A), 095.D-0735(A)\\
NGC 6838 & 072.A-0494(A), 073.D-0211(A), 095.D-0735(A),\\
 & 189.B-0925(A), 189.B-0925(C), C01H, C03H,\\
 & C04H, C10ANS, C118NS, C19H, C216Hr, C52H,\\
 & C53H, H30aH, H5aH, U017Hr, U062NS, U154NS\\
NGC 6864 &  69.B-0305(A), C215D\\
NGC 7006 & C034Hr, C204M, U11H, U21H\\
NGC 7078 & 073.D-0695(A), 080.B-0489(A), 080.B-0784(A),\\
 & 095.D-0539(A), C05H, C11H, C147Hr, C14H,\\
 & C21H, C316Hr, C36H, C53H, C75H, N13H, U09H,\\
 & U27H, U59H\\
NGC 7089 & 084.D-0933(A), 193.B-0936(C), 193.B-0936(F),\\
 & C171D, C174D\\
NGC 7099 & 073.D-0695(A), 085.D-0375(A), 088.B-0403(A),\\
 & 092.D-0477(A), C363Hr, U11H\\
NGC 7492 & C03H, C21H, C88H\\
Pal 2 & U039D\\
Pal 5 & 083.B-0403(A), C50H, U25H\\
Pal 11 & U154NS
\\
Pal 12 & 097.D-0111(B), C11H, C21H, U09H, U47H\\
Pyxis & 089.D-0722(A)\\
Ter 1 & 089.D-0392(A)\\
Ter 3 &  60.A-9700(A),  60.A-9700(G)\\
Ter 5 & 085.D-0377(A), 087.D-0716(A), 087.D-0716(B),\\
 & 087.D-0748(A), 091.D-0115(A), 283.D-5027(A),\\
 & C184D, U060D, U062NS, U074D, U092NS, U097NS,\\
 & U112NS, U146NS, U57NS\\
Ter 6 & 091.D-0115(A)\\
Ter 7 &  67.B-0147(A), 075.D-0075(A), 097.D-0111(B)\\
Ter 8 & 075.D-0075(A), 087.B-0086(A)\\
Ton 2 & 091.D-0389(A)\\
\hline
\end{tabularx}
\end{table}

\clearpage

\section{Sources of surface density profiles, stellar radial velocities and mass functions of individual clusters}

\begin{table}
\caption{Sources of published individual stellar radial velocities (LOS), surface density profiles (SD)
 and stellar mass functions (MF) used in this work in addition to the data published by \citet{baumgardt2017}
  and \citet{sollimabaumgardt2017}.} 
\begin{tabular}[h]{c@{\hspace{0.2cm}}c@{\hspace{0.2cm}}c}
\hline
Name & Source & Type \\ 
\hline
E3       & \citet{vandenberghetal1980} & SD \\[+0.1cm]
ESO 452-SC11 & \citet{bonattobica2008} & SD \\
         & \citet{kochetal2017} & LOS \\
         & \citet{simpsonetal2017} & SD, LOS \\[+0.1cm]
HP 1     & \citet{barbuyetal2016} & LOS  \\[+0.1cm]
IC 1276  & \citet{cote1999} & LOS \\[+0.1cm] 
IC 4499  & \citet{hankeycole2011} & LOS \\
\hline
\end{tabular}
\label{tab:litdata}
\end{table}

\begin{table}
\contcaption{}
\begin{tabular}[h]{c@{\hspace{0.2cm}}c@{\hspace{0.2cm}}c}
\hline
Name & Source & Type \\
\hline
NGC 6205 & \citet{johnsonpilachowski2012} & LOS \\
         & \citet{corderoetal2017} & LOS \\
         & \citet{abolfathietal2017} & LOS \\[+0.1cm]
NGC 6218 & \citet{rastorguevsamus1991} & LOS \\
         & \citet{johnsonpilachowski2006} & LOS \\
         & \citet{abolfathietal2017} & LOS \\[+0.1cm]
NGC 6229 & \citet{johnsonetal2017b} & LOS \\
\hline
\end{tabular}
\end{table}

\begin{table}
\contcaption{}
\begin{tabular}[h]{c@{\hspace{0.2cm}}c@{\hspace{0.2cm}}c}
\hline
Name & Source & Type \\
\hline
NGC 6749 & \citet{canternarosino1981} & SD \\
         & \citet{kaisleretal1997} & SD \\
         & \citet{cote1999} & LOS \\[+0.1cm]
NGC 6752 & \citet{ferraroetal1997} & MF \\
         & \citet{kunderetal2017}  & LOS \\[+0.1cm]
NGC 6760 & \citet{abolfathietal2017} & LOS \\[+0.1cm]
NGC 6809 & \citet{pryoretal1991} & LOS \\
\hline
\end{tabular}
\end{table}

\clearpage

\section{Velocity dispersion profiles of globular clusters calculated from individual stellar radial velocities}

\begin{table}
\caption{Velocity dispersion profiles of globular clusters derived from individual stellar radial velocities. For each bin, the table gives the name of the cluster, the number of stars used to calculate the radial velocity dispersion, the average distance of stars from the cluster centre, and the velocity dispersion together with the $1\sigma$ upper and lower error bars.}
\begin{tabular}{l@{\hspace*{-0.25cm}}rr@{.}lr@{.}l@{\hspace*{-0.1cm}}c@{}c}
\hline
 & \\[-0.3cm]
\multirow{2}{*}{Name} & \multirow{2}{*}{$N_{RV}$} & \multicolumn{2}{c}{$r$} & \multicolumn{2}{c}{$\sigma$} & \multicolumn{1}{c}{$\Delta \sigma_u$} &  \multicolumn{1}{c}{$\Delta \sigma_l$}\\
 & & \multicolumn{2}{c}{[arcsec]} & \multicolumn{2}{c}{[km/sec]} & \multicolumn{1}{c}{[km/sec]} & \multicolumn{1}{c}{[km/sec]}\\[+0.1cm]
\hline
 & \\[-0.3cm]
Arp 2 & 25 & 48 & 40 & 2 & 03 & 0.40 & 0.32 \\
 & 24 & 129 & 77 & 1 & 38 & 0.37 & 0.30 \\
 & \\[-0.3cm]
E 3 & 18 & 124 & 71 & 1 & 56 & 0.47 & 0.39 \\
\hline
\end{tabular}
\end{table}

\addtocounter{table}{-1}
\begin{table}
\caption{(contd.)}
\begin{tabular}{l@{\hspace*{0.05cm}}rr@{.}lr@{.}l@{\hspace*{-0.1cm}}c@{}c}
\hline
 & \\[-0.3cm]
\multirow{2}{*}{Name} & \multirow{2}{*}{$N_{RV}$} & \multicolumn{2}{c}{$r$} & \multicolumn{2}{c}{$\sigma$} & \multicolumn{1}{c}{$\Delta \sigma_u$} &  \multicolumn{1}{c}{$\Delta \sigma_l$}\\
 & & \multicolumn{2}{c}{[arcsec]} & \multicolumn{2}{c}{[km/sec]} & \multicolumn{1}{c}{[km/sec]} & \multicolumn{1}{c}{[km/sec]}\\[+0.1cm]
\hline
 & 65 & 193 & 22 & 4 & 94 & 0.48 & 0.42 \\
 & 65 & 247 & 27 & 3 & 38 & 0.33 & 0.29 \\
 & 65 & 324 & 90 & 3 & 92 & 0.38 & 0.32 \\
 & 55 & 464 & 73 & 3 & 37 & 0.37 & 0.31 \\
 & \\
\hline
\end{tabular}
\end{table}

\addtocounter{table}{-1}
\begin{table}
\caption{(contd.)}
\begin{tabular}{l@{\hspace*{0.05cm}}rr@{.}lr@{.}l@{\hspace*{-0.1cm}}c@{}c}
\hline
 & \\[-0.3cm]
\multirow{2}{*}{Name} & \multirow{2}{*}{$N_{RV}$} & \multicolumn{2}{c}{$r$} & \multicolumn{2}{c}{$\sigma$} & \multicolumn{1}{c}{$\Delta \sigma_u$} &  \multicolumn{1}{c}{$\Delta \sigma_l$}\\
 & & \multicolumn{2}{c}{[arcsec]} & \multicolumn{2}{c}{[km/sec]} & \multicolumn{1}{c}{[km/sec]} & \multicolumn{1}{c}{[km/sec]}\\[+0.1cm]
\hline
 & 125 & 90 & 48 & 17 & 05 & 1.14 & 1.03 \\
 & 250 & 1056 & 53 & 8 & 40 & 0.39 & 0.37 \\
 & 250 & 121 & 52 & 17 & 87 & 0.83 & 0.78 \\
 & 250 & 1459 & 37 & 7 & 66 & 0.36 & 0.34 \\
 & 250 & 154 & 90 & 15 & 69 & 0.73 & 0.68 \\
\hline
\end{tabular}
\end{table}

\addtocounter{table}{-1}
\begin{table}
\caption{(contd.)}
\begin{tabular}{l@{\hspace*{0.05cm}}rr@{.}lr@{.}l@{\hspace*{-0.1cm}}c@{}c}
\hline
 & \\[-0.3cm]
\multirow{2}{*}{Name} & \multirow{2}{*}{$N_{RV}$} & \multicolumn{2}{c}{$r$} & \multicolumn{2}{c}{$\sigma$} & \multicolumn{1}{c}{$\Delta \sigma_u$} &  \multicolumn{1}{c}{$\Delta \sigma_l$}\\
 & & \multicolumn{2}{c}{[arcsec]} & \multicolumn{2}{c}{[km/sec]} & \multicolumn{1}{c}{[km/sec]} & \multicolumn{1}{c}{[km/sec]}\\[+0.1cm]
\hline
 & 60 & 111 & 40 & 5 & 46 & 0.55 & 0.47 \\
 & 60 & 160 & 62 & 5 & 38 & 0.55 & 0.47 \\
 & 60 & 229 & 38 & 4 & 75 & 0.49 & 0.42 \\
 & 60 & 330 & 78 & 4 & 19 & 0.43 & 0.37 \\
 & 20 & 453 & 04 & 3 & 68 & 0.70 & 0.55 \\
\hline
\end{tabular}
\end{table}

\addtocounter{table}{-1}
\begin{table}
\caption{(contd.)}
\begin{tabular}{l@{\hspace*{0.05cm}}rr@{.}lr@{.}l@{\hspace*{-0.1cm}}c@{}c}
\hline
 & \\[-0.3cm]
\multirow{2}{*}{Name} & \multirow{2}{*}{$N_{RV}$} & \multicolumn{2}{c}{$r$} & \multicolumn{2}{c}{$\sigma$} & \multicolumn{1}{c}{$\Delta \sigma_u$} &  \multicolumn{1}{c}{$\Delta \sigma_l$}\\
 & & \multicolumn{2}{c}{[arcsec]} & \multicolumn{2}{c}{[km/sec]} & \multicolumn{1}{c}{[km/sec]} & \multicolumn{1}{c}{[km/sec]}\\[+0.1cm]
\hline
 & \\[-0.3cm]
NGC 6273 & 65 & 46 & 99 & 10 & 28 & 0.98 & 0.86 \\
 & 65 & 84 & 92 & 8 & 94 & 0.86 & 0.74 \\
 & 65 & 135 & 85 & 8 & 79 & 0.85 & 0.73 \\
 & 65 & 224 & 75 & 6 & 75 & 0.65 & 0.56 \\
\hline
\end{tabular}
\end{table}

\addtocounter{table}{-1}
\begin{table}
\caption{(contd.)}
\begin{tabular}{l@{\hspace*{0.05cm}}rr@{.}lr@{.}l@{\hspace*{-0.1cm}}c@{}c}
\hline
 & \\[-0.3cm]
\multirow{2}{*}{Name} & \multirow{2}{*}{$N_{RV}$} & \multicolumn{2}{c}{$r$} & \multicolumn{2}{c}{$\sigma$} & \multicolumn{1}{c}{$\Delta \sigma_u$} &  \multicolumn{1}{c}{$\Delta \sigma_l$}\\
 & & \multicolumn{2}{c}{[arcsec]} & \multicolumn{2}{c}{[km/sec]} & \multicolumn{1}{c}{[km/sec]} & \multicolumn{1}{c}{[km/sec]}\\[+0.1cm]
\hline
 & 56 & 329 & 81 & 5 & 82 & 0.61 & 0.53 \\
 & \\[-0.3cm]
NGC 6440 & 35 & 21 & 47 & 11 & 66 & 1.56 & 1.29 \\
 & 35 & 66 & 99 & 9 & 62 & 1.29 & 1.06 \\
 & 32 & 141 & 17 & 7 & 54 & 1.13 & 0.92 \\
\hline
\end{tabular}
\end{table}

\addtocounter{table}{-1}
\begin{table}
\caption{(contd.)}
\begin{tabular}{l@{\hspace*{0.05cm}}rr@{.}lr@{.}l@{\hspace*{-0.1cm}}c@{}c}
\hline
 & \\[-0.3cm]
\multirow{2}{*}{Name} & \multirow{2}{*}{$N_{RV}$} & \multicolumn{2}{c}{$r$} & \multicolumn{2}{c}{$\sigma$} & \multicolumn{1}{c}{$\Delta \sigma_u$} &  \multicolumn{1}{c}{$\Delta \sigma_l$}\\
 & & \multicolumn{2}{c}{[arcsec]} & \multicolumn{2}{c}{[km/sec]} & \multicolumn{1}{c}{[km/sec]} & \multicolumn{1}{c}{[km/sec]}\\[+0.1cm]
\hline
 & \\[-0.3cm]
NGC 6715 & 65 & 34 & 65 & 8 & 57 & 0.82 & 0.71 \\
 & 65 & 61 & 60 & 8 & 51 & 0.81 & 0.70 \\
 & 65 & 93 & 75 & 7 & 99 & 0.76 & 0.66 \\
 & 65 & 125 & 34 & 8 & 17 & 0.78 & 0.68 \\
\hline
\end{tabular}
\end{table}

\addtocounter{table}{-1}
\begin{table}
\caption{(contd.)}
\begin{tabular}{l@{\hspace*{0.05cm}}rr@{.}lr@{.}l@{\hspace*{-0.1cm}}c@{}c}
\hline
 & \\[-0.3cm]
\multirow{2}{*}{Name} & \multirow{2}{*}{$N_{RV}$} & \multicolumn{2}{c}{$r$} & \multicolumn{2}{c}{$\sigma$} & \multicolumn{1}{c}{$\Delta \sigma_u$} &  \multicolumn{1}{c}{$\Delta \sigma_l$}\\
 & & \multicolumn{2}{c}{[arcsec]} & \multicolumn{2}{c}{[km/sec]} & \multicolumn{1}{c}{[km/sec]} & \multicolumn{1}{c}{[km/sec]}\\[+0.1cm]
\hline
 & \\[-0.3cm]
NGC 7089 & 50 & 38 & 48 & 9 & 52 & 1.04 & 0.89 \\
 & 50 & 76 & 25 & 8 & 07 & 0.89 & 0.76 \\
 & 50 & 115 & 04 & 8 & 15 & 0.90 & 0.76 \\
 & 50 & 148 & 38 & 7 & 00 & 0.78 & 0.66 \\
\hline
\end{tabular}
\label{veldistab}
\end{table}

\clearpage

\section{Individual stellar radial velocities of stars in the fields of 53 globular clusters}

\begin{table*}
\caption{Individual stellar radial velocities for stars in the field of Arp 2. The table gives the cluster name, the 2MASS ID, the right ascension and declination, the average heliocentric radial velocity and its 1$\sigma$ error, the distance from the cluster centre, the 2MASS $J$ and $K_S$ band magnitudes, the membership probability based on the radial velocity and the number of radial velocity measurements. For stars with multiple radial velocity measurements, the probability that the star has a constant radial velocity is given in the final column. Full versions of Tables \ref{indveltabstart} to \ref{indveltabend} are only available online.}
\begin{tabular}{l@{$\;\;\;$}c@{$\;\;\;$}c@{$\;\;\;$}c@{$\;\;\;$}r@{$\; \pm \;$}lrcc@{$\;\;\;$}c@{$\;\;\;$}c@{$\;\;\;$}l}
\hline
 & \\[-0.3cm]
\multirow{2}{*}{Name} & \multirow{2}{*}{2MASS ID} & RA & DEC & \multicolumn{2}{c}{$R_V$} & \multicolumn{1}{c}{$d$} & $J$ & $K_S$ & Prob. &  \multirow{2}{*}{$N_{RV}$} & Prob.\\ 
 & & [J2000] & [J2000] & \multicolumn{2}{c}{[km/sec]} & \multicolumn{1}{c}{['']} & [mag] & [mag] & Mem. &  & Single\\  
\hline
 & \\[-0.3cm]
Arp 2 &                  &  292.055125 &  -30.408778 &  121.22 &   2.75 & 443.07  &    &    & 0.704 & 1 &  \\
Arp 2 & 19281376-3020014 &  292.057344 &  -30.333748 &  153.94 &   0.88 & 400.72  &  14.74 $\pm$   0.04 &  13.97 $\pm$   0.07 & 0.000 & 1 &  \\
Arp 2 & 19281572-3016559 &  292.065539 &  -30.282202 &  -61.18 &   1.64 & 452.74  &  15.25 $\pm$   0.05 &  14.80 $\pm$   0.12 & 0.000 & 1 &  \\
Arp 2 & 19283875-3020200 &  292.161484 &  -30.338905 &  125.30 &   0.62 &  91.86  &  16.69 $\pm$   0.15 &  17.00\hspace*{0.95cm}  & 0.129 & 1 &  \\
Arp 2 & 19284024-3021598 &  292.167701 &  -30.366631 &  123.27 &   0.42 &  63.79  &  15.50 $\pm$   0.07 &  14.89 $\pm$   0.14 & 0.669 & 2 & 0.893\\
 \multicolumn{1}{c}{$\vdots$}  & $\vdots$  & $\vdots$ & $\vdots$ &\multicolumn{2}{c}{$\vdots$} & \multicolumn{1}{c}{$\vdots$} & $\vdots$ & $\vdots$ & $\vdots$ & $\vdots$ \\
\hline
\end{tabular}
\label{indveltabstart}
\end{table*}

\setcounter{table}{52}
\begin{table*}
\caption{Same as Table \ref{indveltabstart} for stars in the field of Ter 8.}
\begin{tabular}{l@{$\;\;\;$}c@{$\;\;\;$}c@{$\;\;\;$}c@{$\;\;\;$}r@{$\; \pm \;$}lrcc@{$\;\;\;$}c@{$\;\;\;$}c@{$\;\;\;$}l}
\hline
 & \\[-0.3cm]
\multirow{2}{*}{Name} & \multirow{2}{*}{2MASS ID} & RA & DEC & \multicolumn{2}{c}{$R_V$} & \multicolumn{1}{c}{$d$} & $J$ & $K_S$ & Prob. & \multirow{2}{*}{$N_{RV}$} & Prob.\\ 
 & & [J2000] & [J2000] & \multicolumn{2}{c}{[km/sec]} & \multicolumn{1}{c}{['']} & [mag] & [mag] & Mem. & & Single\\  
\hline
 & \\[-0.3cm]
Ter 8 &                  &  295.286292 &  -34.052194 &  145.71 &   2.25 & 482.66  &    &    & 0.261 & 1 &  \\
Ter 8 &                  &  295.302250 &  -34.012500 &  147.82 &   2.29 & 399.02  &    &    & 0.779 & 1 &  \\
Ter 8 & 19411316-3403599 &  295.304859 &  -34.066650 &  -66.97 &   0.99 & 457.49  &  15.65 $\pm$   0.06 &  15.70 $\pm$   0.23 & 0.000 & 1 &  \\
Ter 8 & 19411449-3405234 &  295.310392 &  -34.089840 &  -48.27 &   2.67 & 494.02  &  16.73 $\pm$   0.14 &  16.00\hspace*{0.95cm}  & 0.000 & 1 &  \\
Ter 8 &                  &  295.316917 &  -33.947944 &  164.88 &   2.75 & 398.44  &    &    & 0.000 & 1 &  \\
\multicolumn{1}{c}{$\vdots$}  & $\vdots$  & $\vdots$ & $\vdots$ &\multicolumn{2}{c}{$\vdots$} & \multicolumn{1}{c}{$\vdots$} & $\vdots$ & $\vdots$ & $\vdots$ & $\vdots$ \\
\hline
\end{tabular}
\label{indveltabend}
\end{table*}

\clearpage

\section{Fits of the surface density and velocity dispersion profiles of individual clusters}

\begin{figure*}
\begin{center}
\includegraphics[width=17cm]{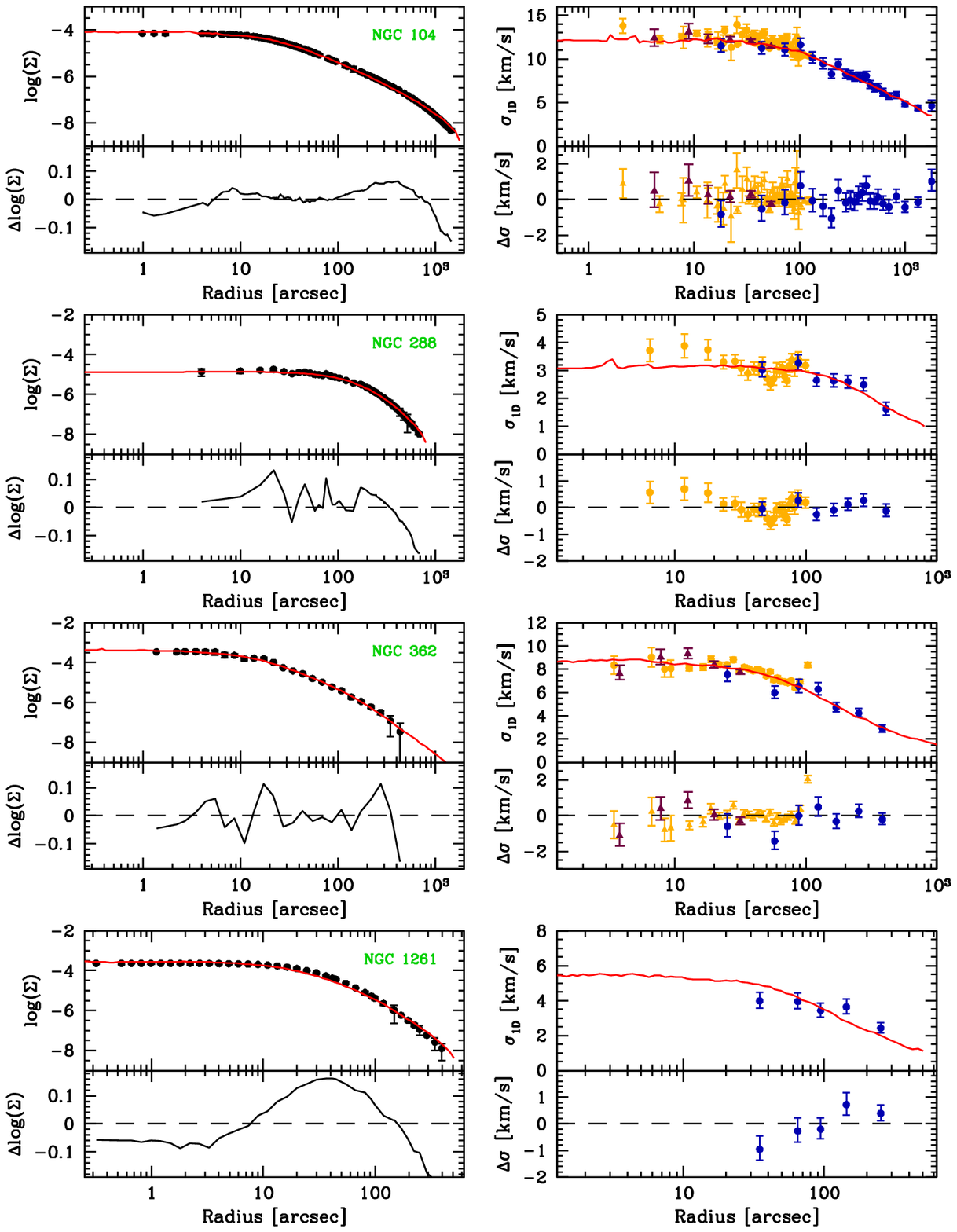}
\end{center}
\vspace*{-0.5cm}
\caption{Fit of the surface density profiles (left panels) and velocity dispersion profiles (right panels) for NGC 104,
NGC 288, NGC 362 and NGC 1261. In the right panels, the observed proper motion dispersion profile
of \citet{watkinsetal2015a} is shown by orange circles while the radial velocity dispersion profile derived in this work is shown 
by blue circles. In order to convert proper motions to velocities we use the distances given in Table~2. Triangles mark the radial velocity dispersion profiles from \citet{kamannetal2018}. Red curves show the surface density (left panel) and line-of-sight velocity dispersion (right panel) of the best-fitting $N$-body model
for each cluster. The lower panels show the differences between
the observed data and the $N$-body models. The $N$-body data usually provides an excellent fit to the observed data for the depicted clusters.}
\label{fig1a}
\end{figure*}

\begin{figure*}
\begin{center}
\includegraphics[width=17cm]{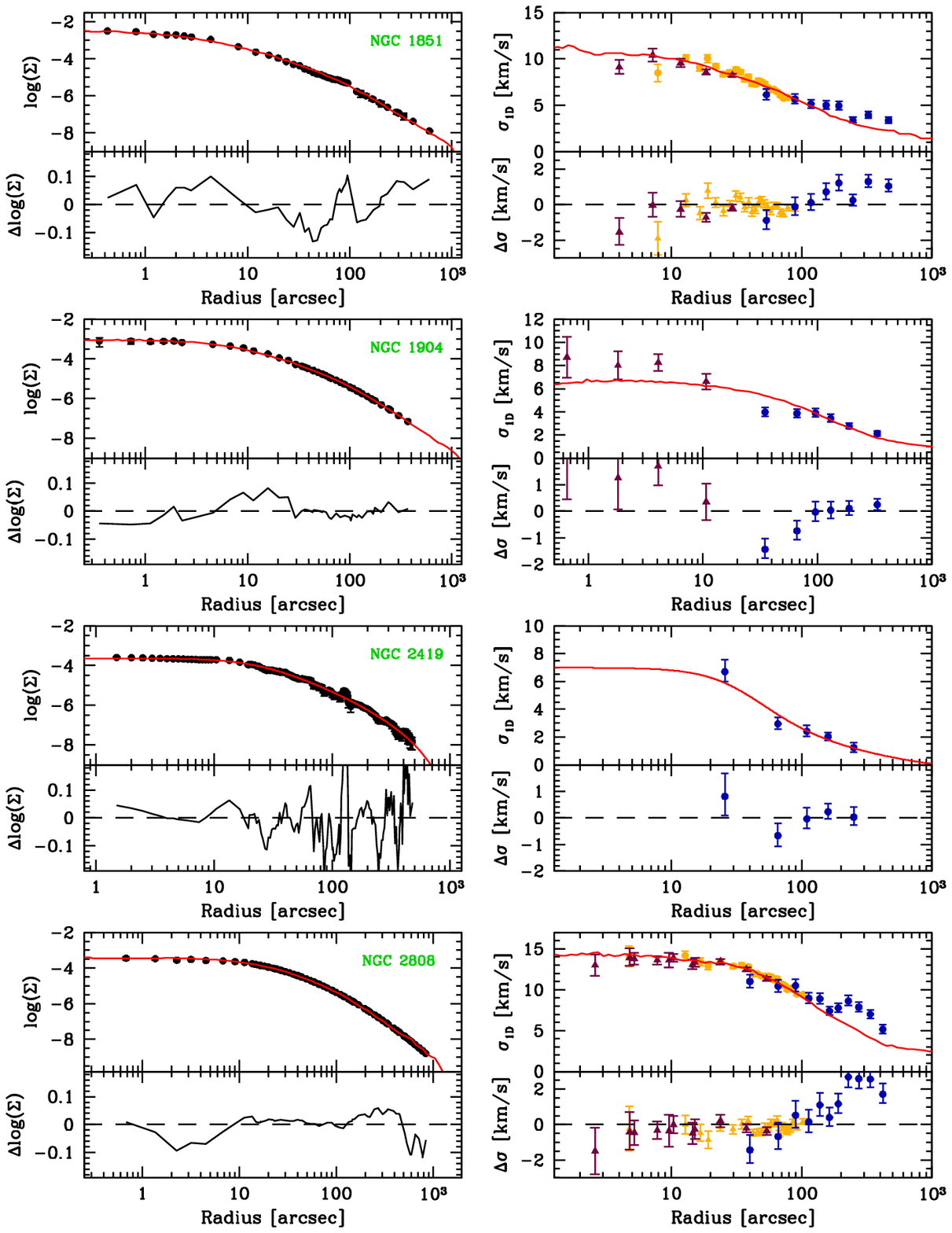}
\end{center}
\caption{Same as Fig. \ref{fig1a} for NGC 1851, NGC 1904, NGC 2419 and NGC 2808. As discussed in the main text, we use a radially
anisotropic King model to fit NGC 2419.\hspace*{7cm}}
\label{fig2a}
\end{figure*}

Figs. \ref{fig1a} to \ref{fig15a} depict our fits to the observed surface density and velocity dispersion profiles for clusters
with more than 100 member stars.
The surface densities in the left panels are normalized to~1. In the right panels, the proper motion data 
from \citet{watkinsetal2015a}
are shown by orange circles while the radial velocity dispersion profiles derived in this work are shown by blue circles.
The predictions of the best-fitting $N$-body models are shown as solid, red lines. For clarity we show only the predicted
radial velocity dispersion profiles. The proper motion dispersion profiles usually agree with the radial velocity ones
to within a few \%.

\begin{figure*}
\begin{center}
\includegraphics[width=17cm]{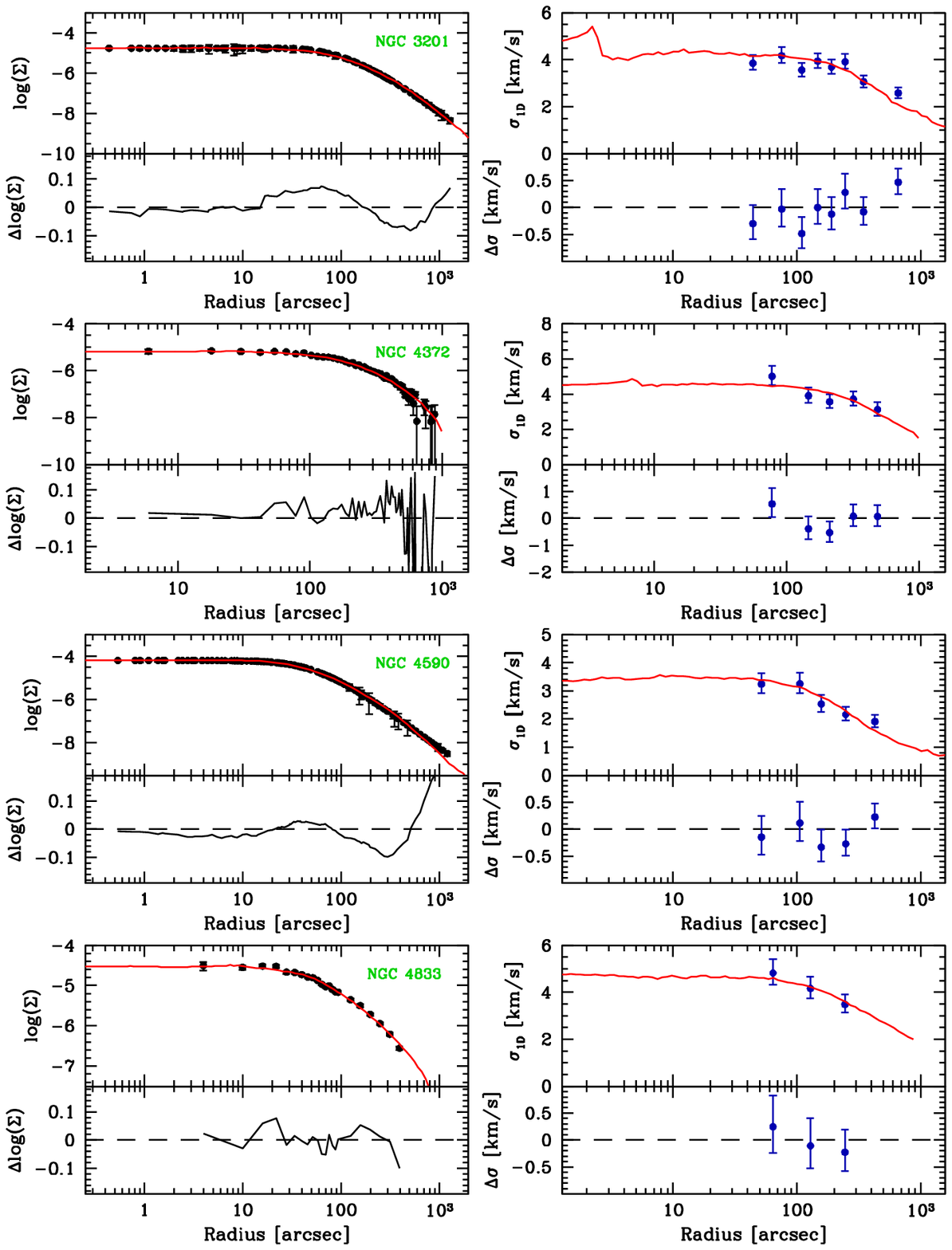}
\end{center}
\caption{Same as Fig. \ref{fig1a} for NGC 3201, NGC 4372, NGC 4590 and NGC 4833.\hspace*{7cm}}
\end{figure*}

\begin{figure*}
\begin{center}
\includegraphics[width=17cm]{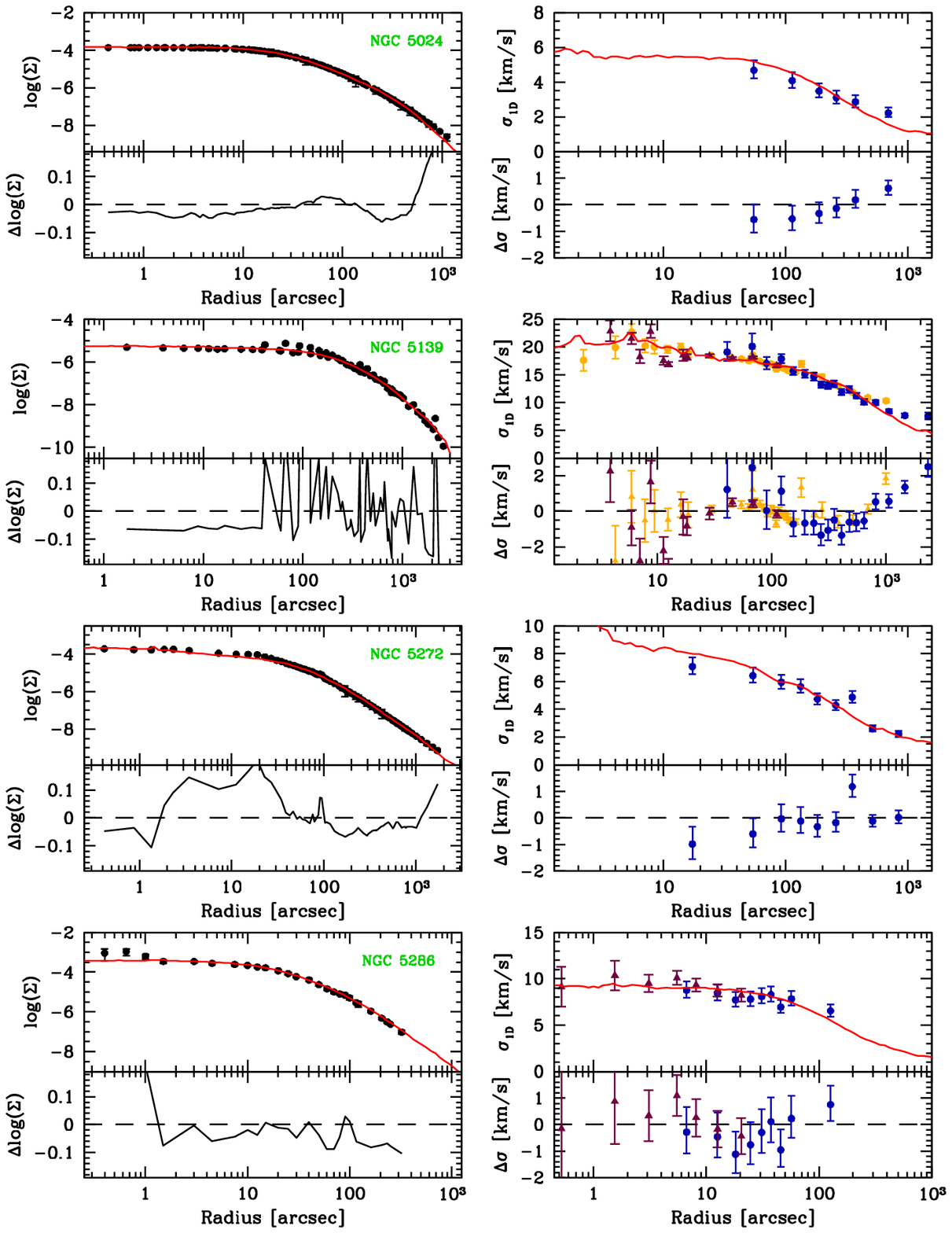}
\end{center}
\caption{Same as Fig. \ref{fig1a} for NGC 5024, NGC 5139, NGC 5272 and NGC 5286. The red, solid lines for NGC 5139 show the best-fitting
no IMBH model from \citet{baumgardtetal2018}.\hspace*{2cm}}
\end{figure*}

\begin{figure*}
\begin{center}
\includegraphics[width=17cm]{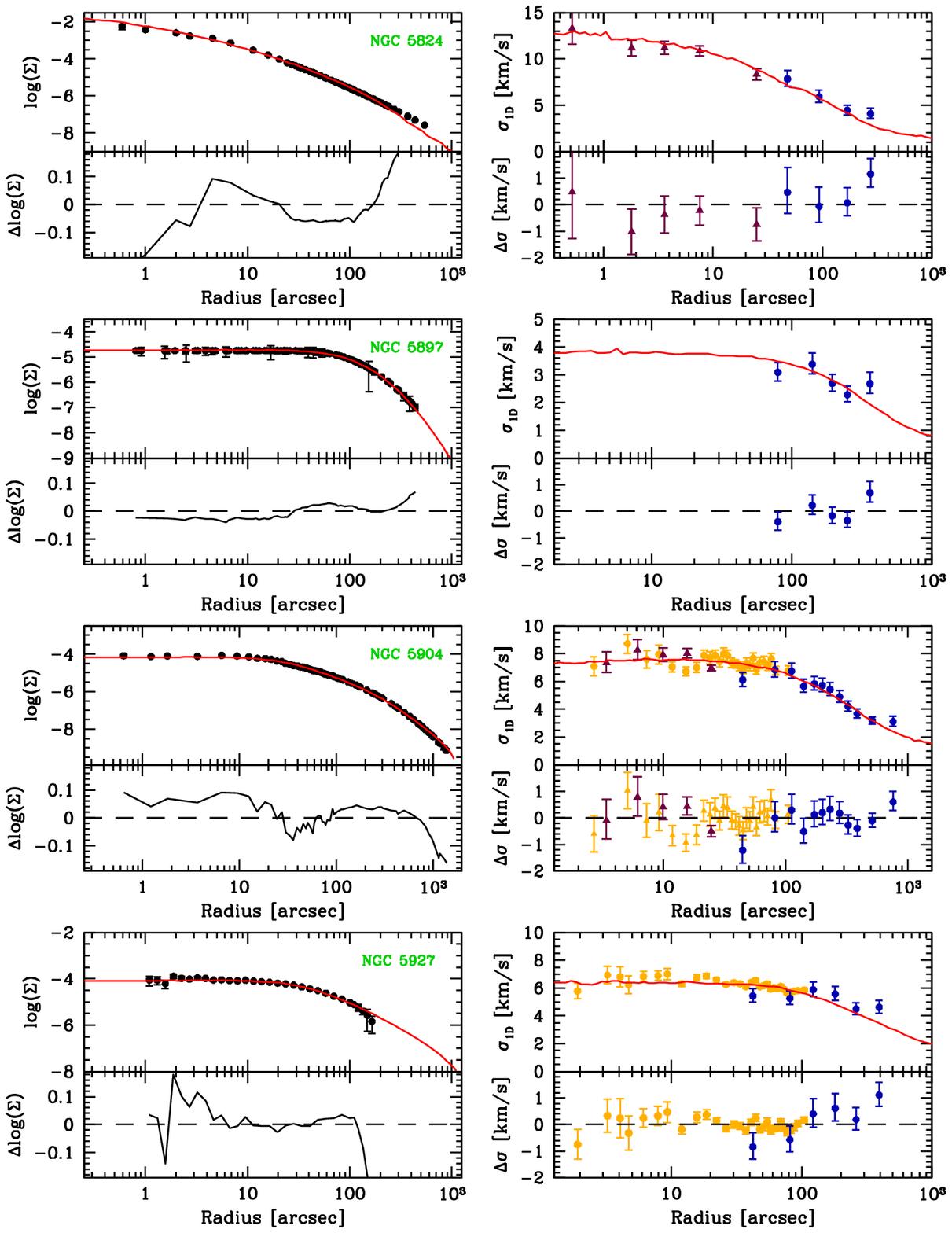}
\end{center}
\caption{Same as Fig. \ref{fig1a} for NGC 5824, NGC 5897, NGC 5904 and NGC 5927.\hspace*{7cm}}
\end{figure*}

\begin{figure*}
\begin{center}
\includegraphics[width=17cm]{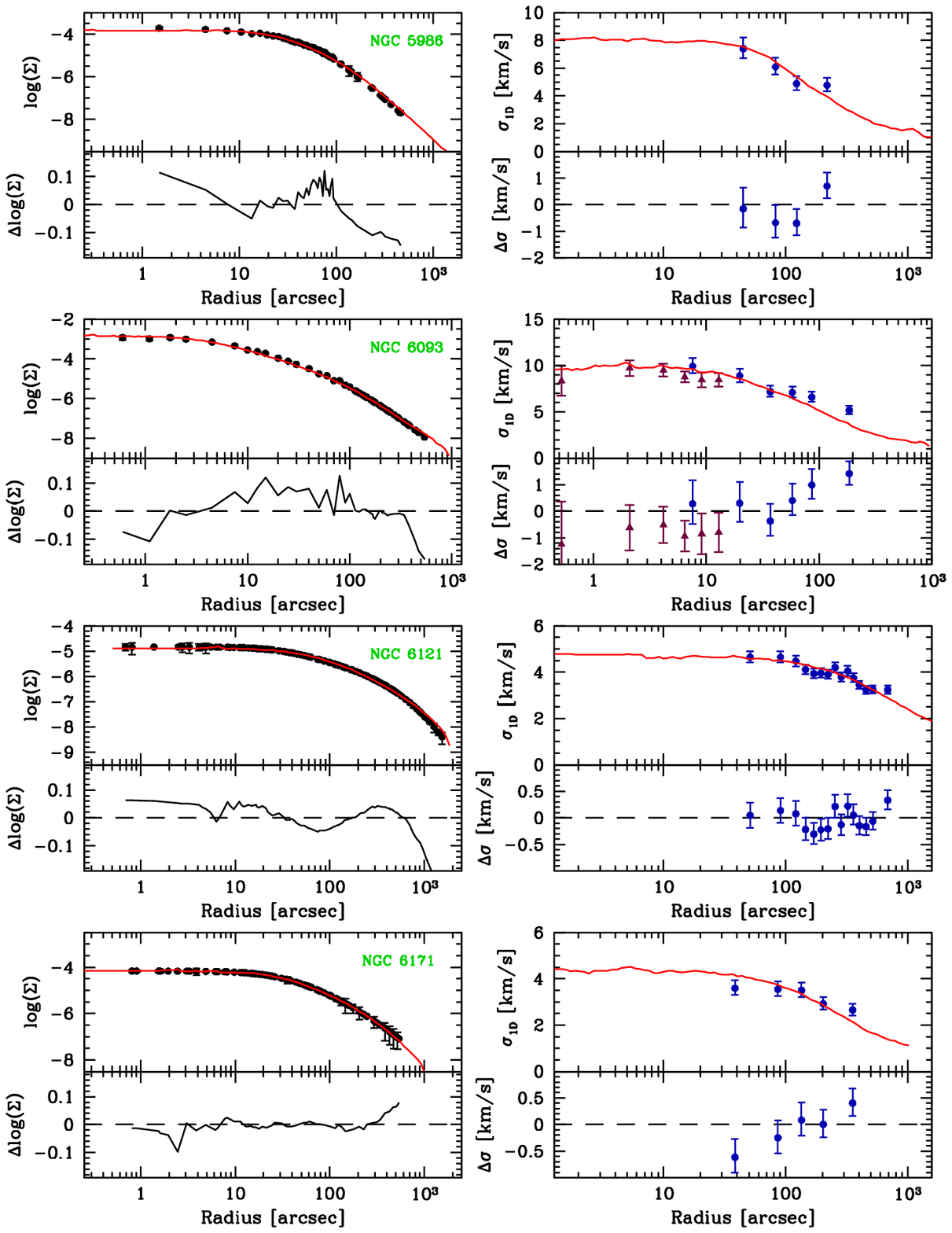}
\end{center}
\caption{Same as Fig. \ref{fig1a} for NGC 5986, NGC 6093, NGC 6121 and NGC 6171.\hspace*{7cm}}
\end{figure*}

\begin{figure*}
\begin{center}
\includegraphics[width=17cm]{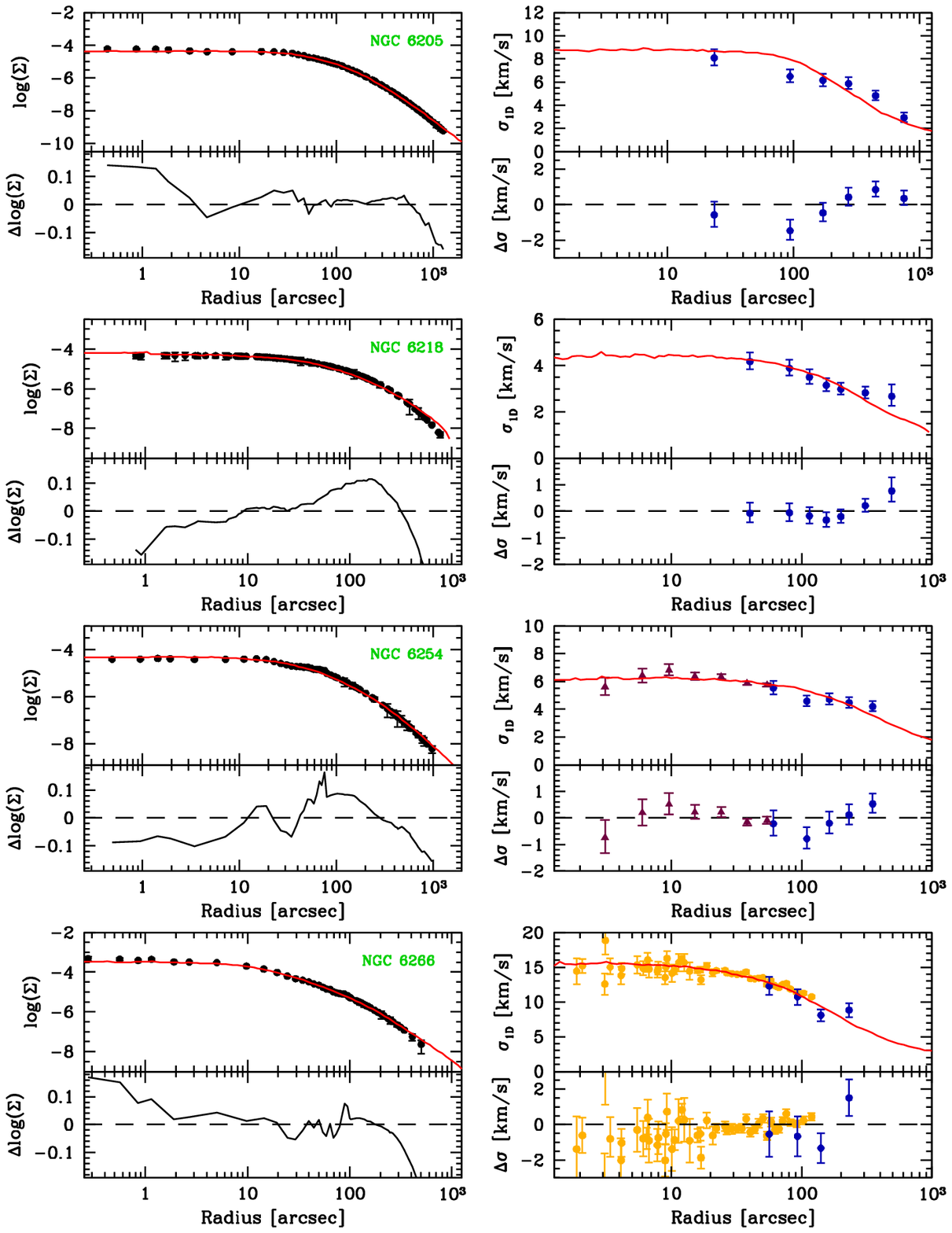}
\end{center}
\caption{Same as Fig. \ref{fig1a} for NGC 6205, NGC 6218, NGC 6254, and NGC 6266.\hspace*{7cm}}
\end{figure*}

\begin{figure*}
\begin{center}
\includegraphics[width=17cm]{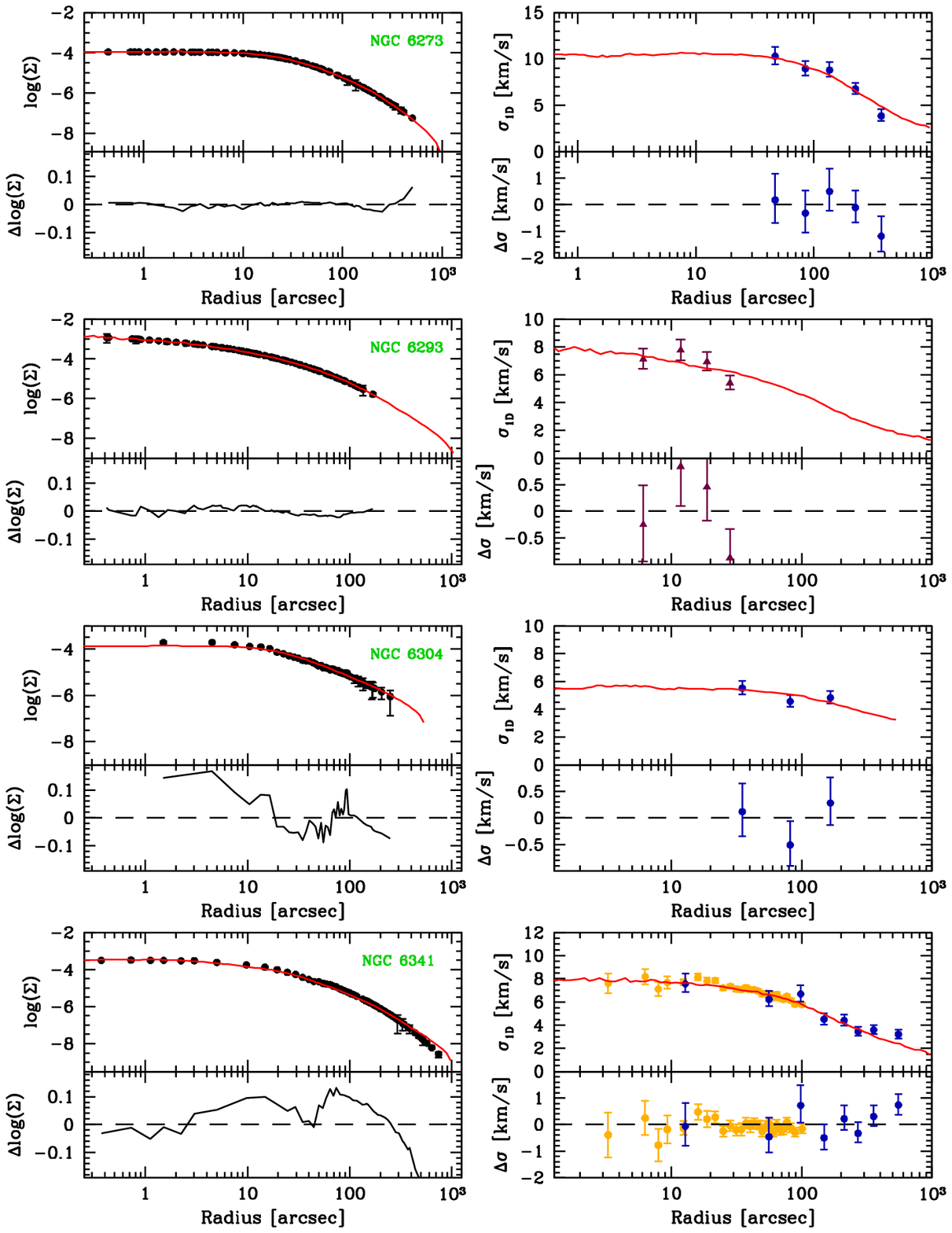}
\end{center}
\caption{Same as Fig. \ref{fig1a} for NGC 6273, NGC 6293, NGC 6304 and NGC 6341.\hspace*{7cm}}
\end{figure*}

\begin{figure*}
\begin{center}
\includegraphics[width=17cm]{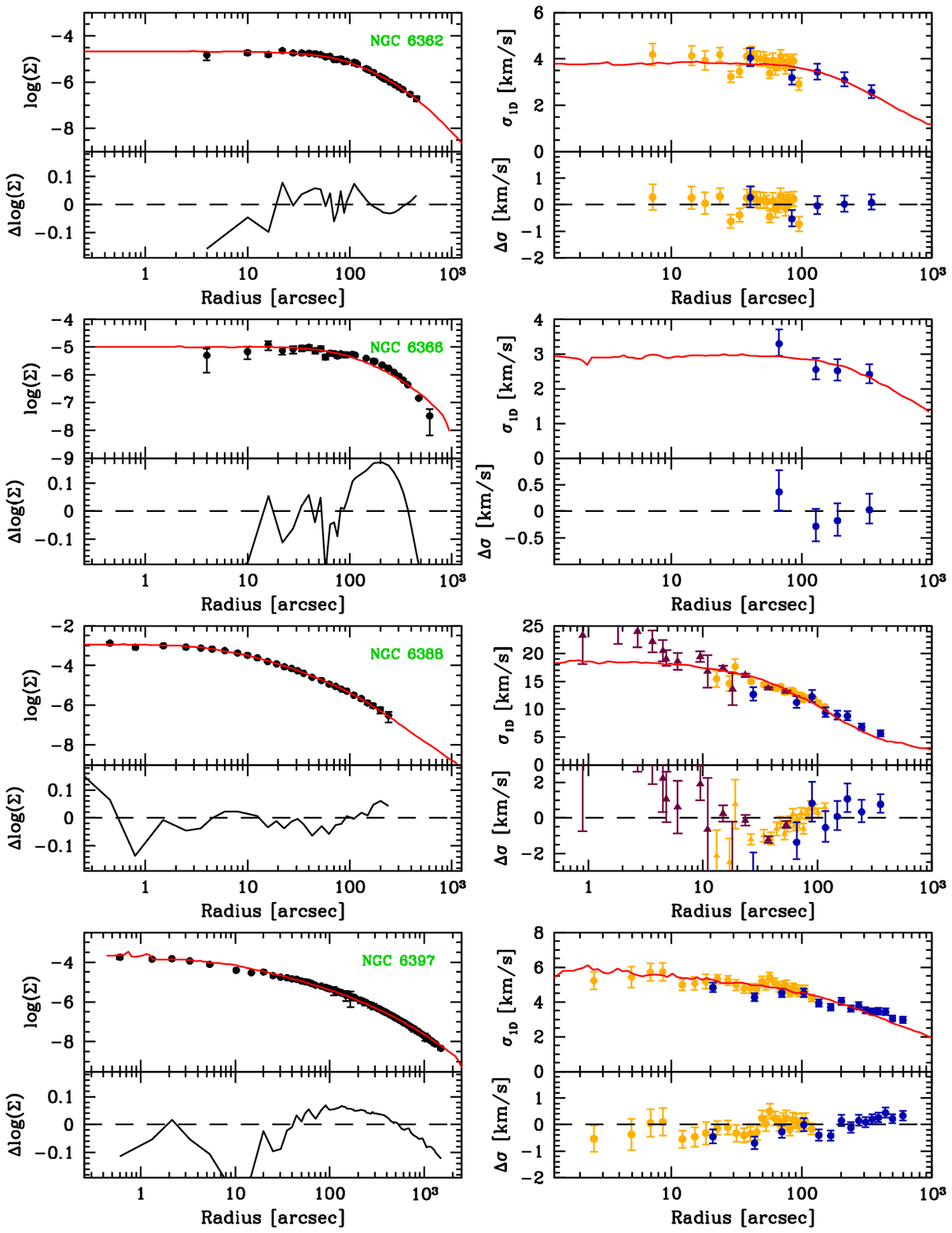}
\end{center}
\caption{Same as Fig. \ref{fig1a} for NGC 6362, NGC 6366, NGC 6388 and NGC 6397.\hspace*{7cm}}
\end{figure*}

\begin{figure*}
\begin{center}
\includegraphics[width=17cm]{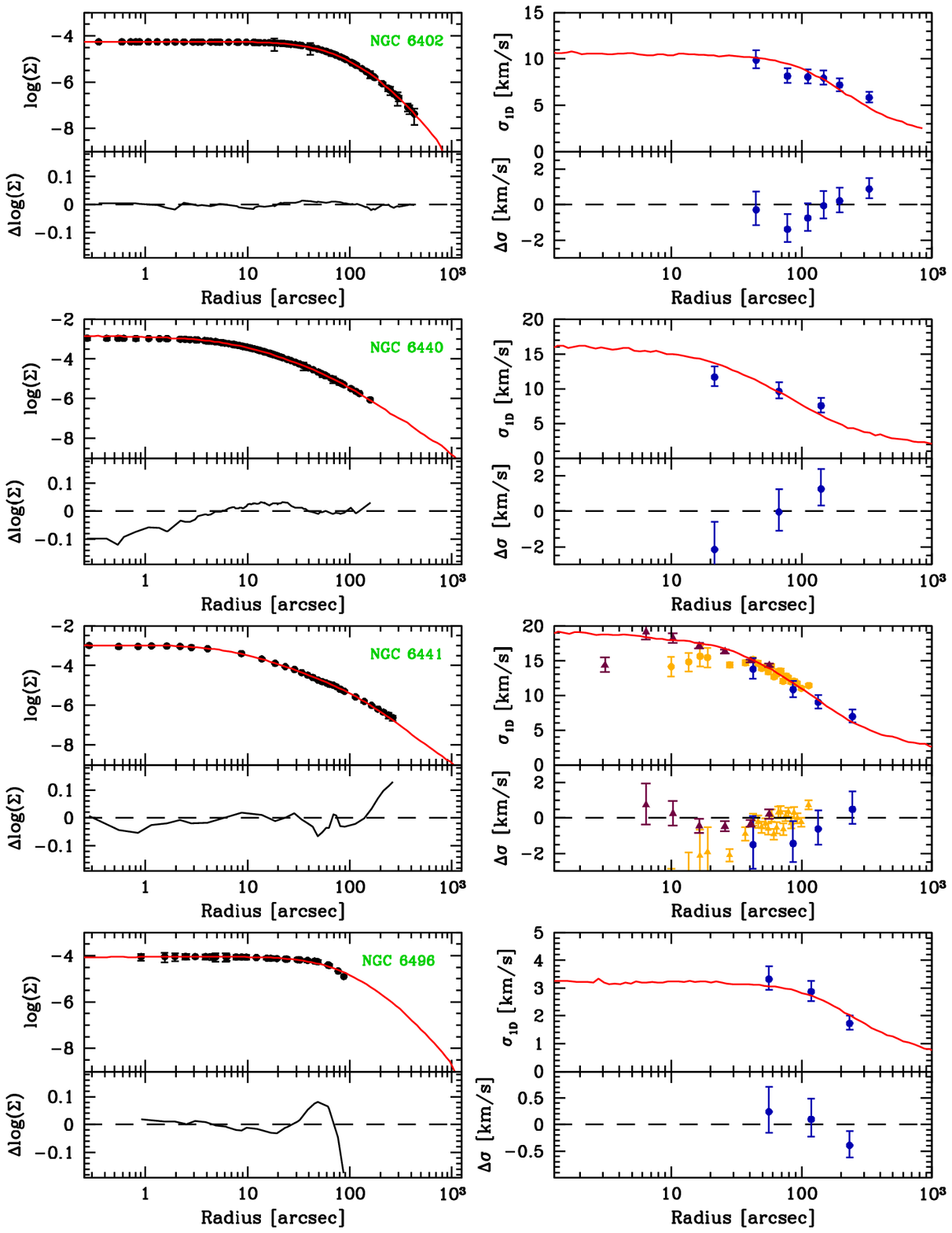}
\end{center}
\caption{Same as Fig. \ref{fig1a} for NGC 6402, NGC 6440, NGC 6441 and NGC 6496.\hspace*{7cm}}
\end{figure*}

\begin{figure*}
\begin{center}
\includegraphics[width=17cm]{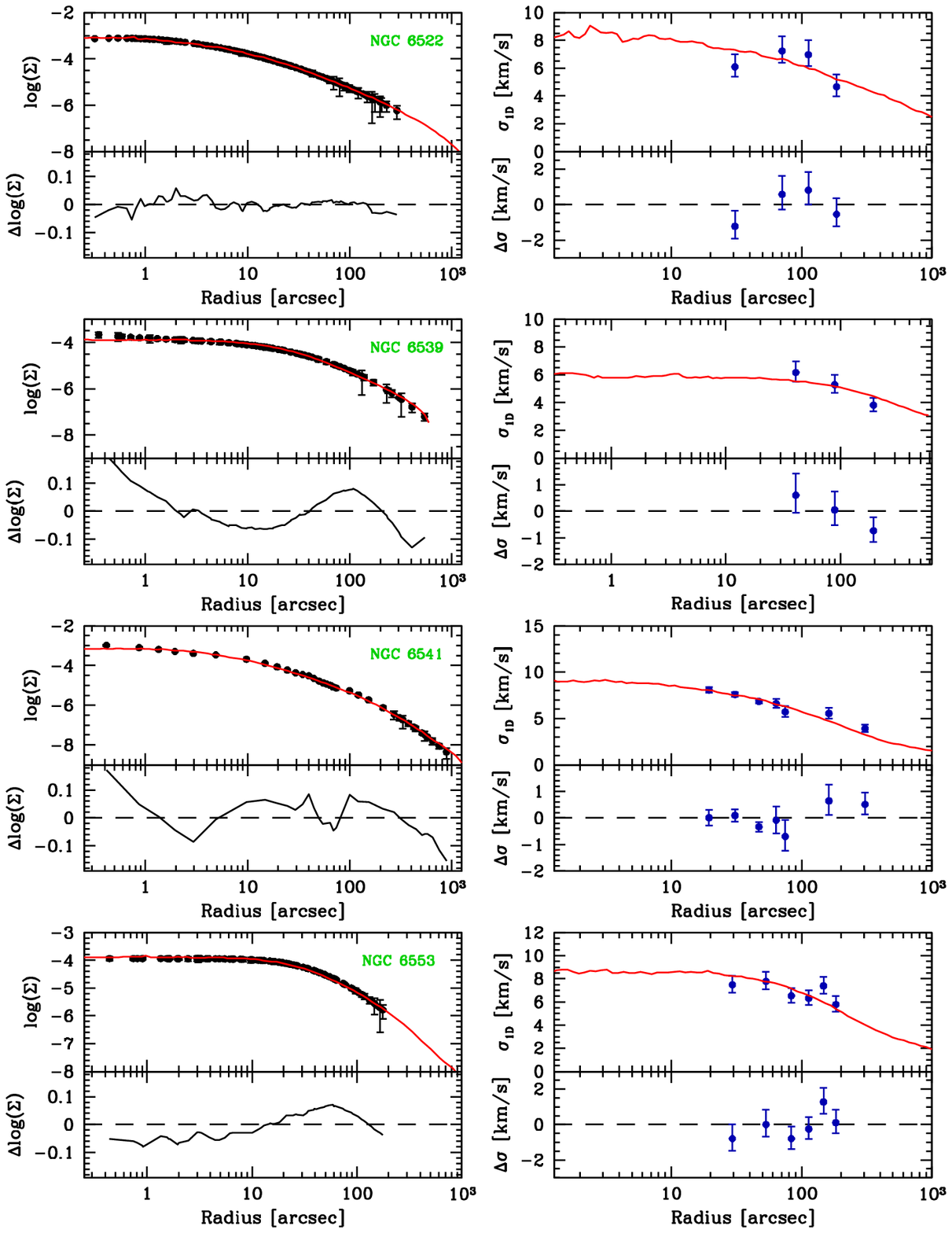}
\end{center}
\caption{Same as Fig. \ref{fig1a} for NGC 6522, NGC 6539, NGC 6541 and NGC 6553.\hspace*{7cm}}
\end{figure*}

\begin{figure*}
\begin{center}
\includegraphics[width=17cm]{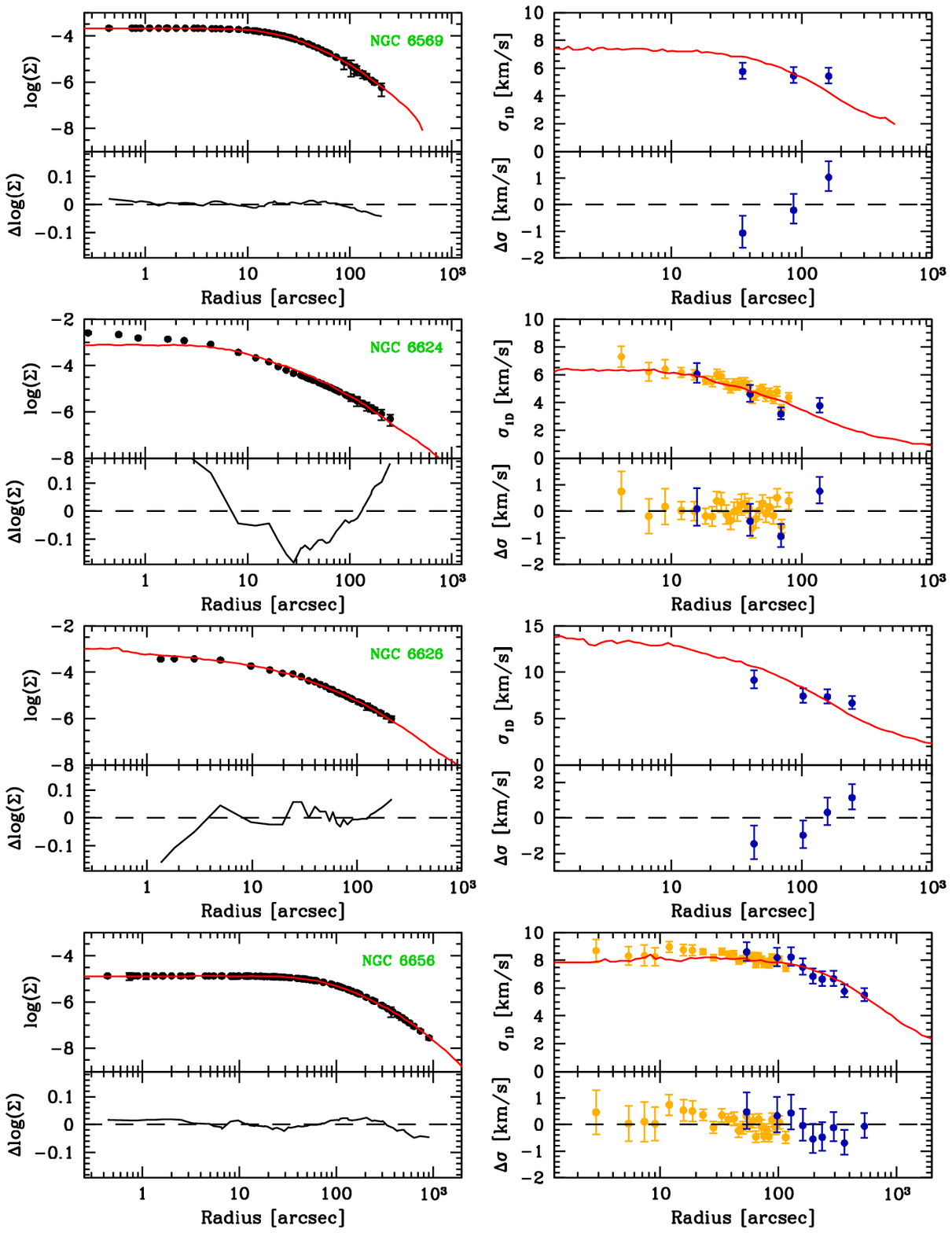}
\end{center}
\caption{Same as Fig. \ref{fig1a} for NGC 6569, NGC 6624, NGC 6626 and NGC 6656.\hspace*{7cm}}
\label{fig12a}
\end{figure*}

\begin{figure*}
\begin{center}
\includegraphics[width=17cm]{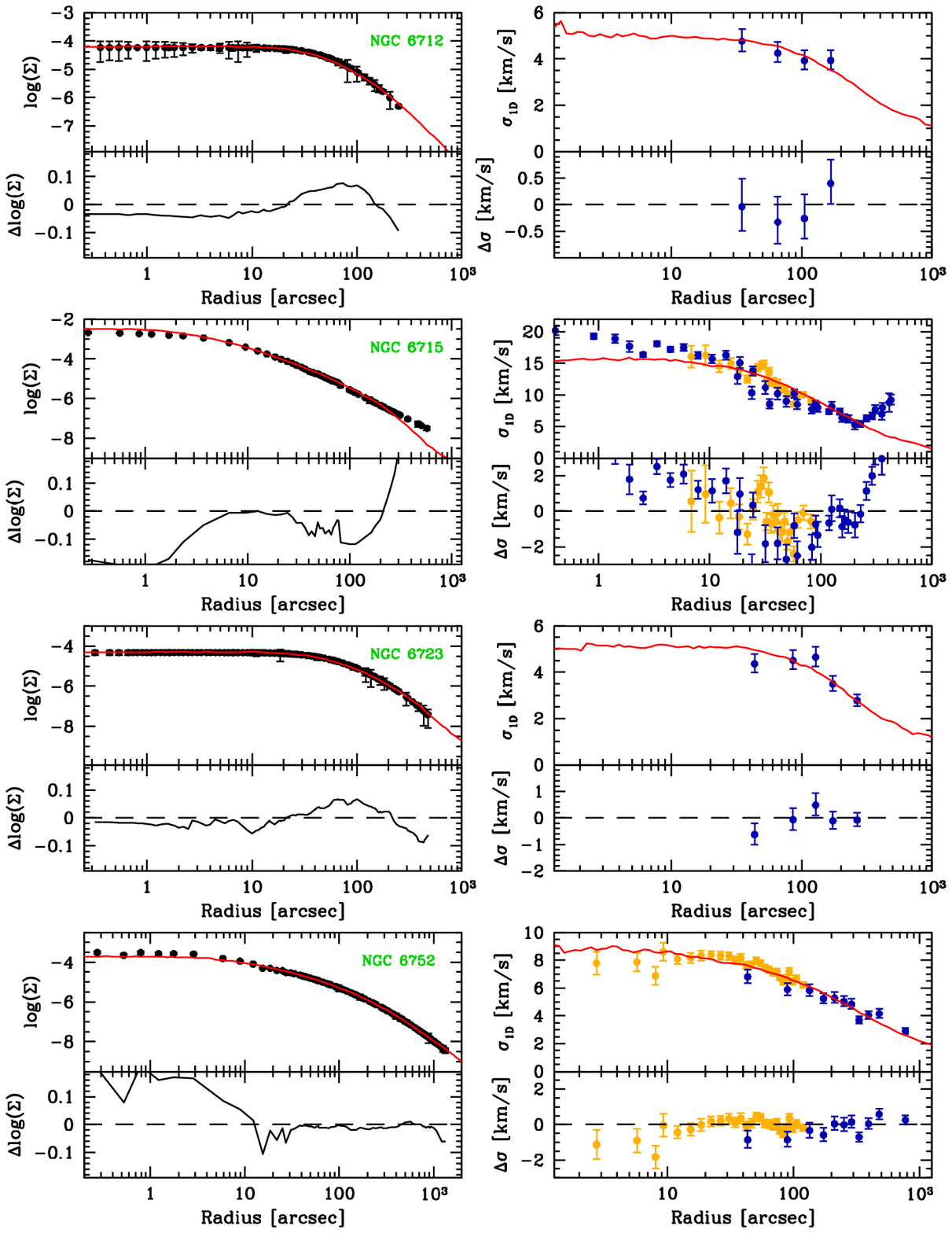}
\end{center}
\caption{Same as Fig. \ref{fig1a} for NGC 6712, NGC 6715, NGC 6723 and NGC 6752.\hspace*{7cm}}
\label{fig13a}
\end{figure*}

\begin{figure*}
\begin{center}
\includegraphics[width=17cm]{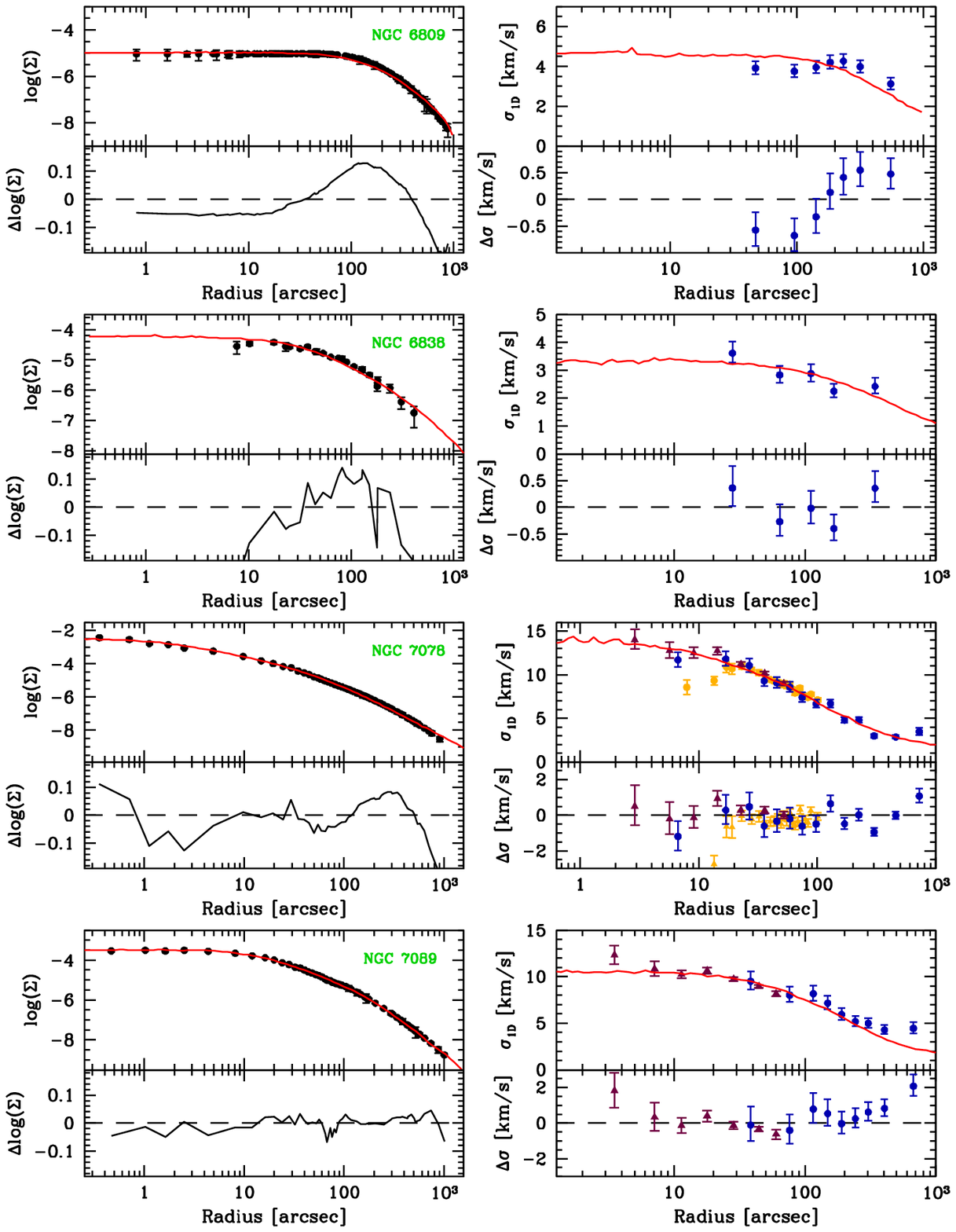}
\end{center}
\caption{Same as Fig. \ref{fig1a} for NGC 6809, NGC 6838, NGC 7078 and NGC 7089.\hspace*{7cm}}
\label{fig14a}
\end{figure*}

\begin{figure*}
\begin{center}
\includegraphics[width=17cm]{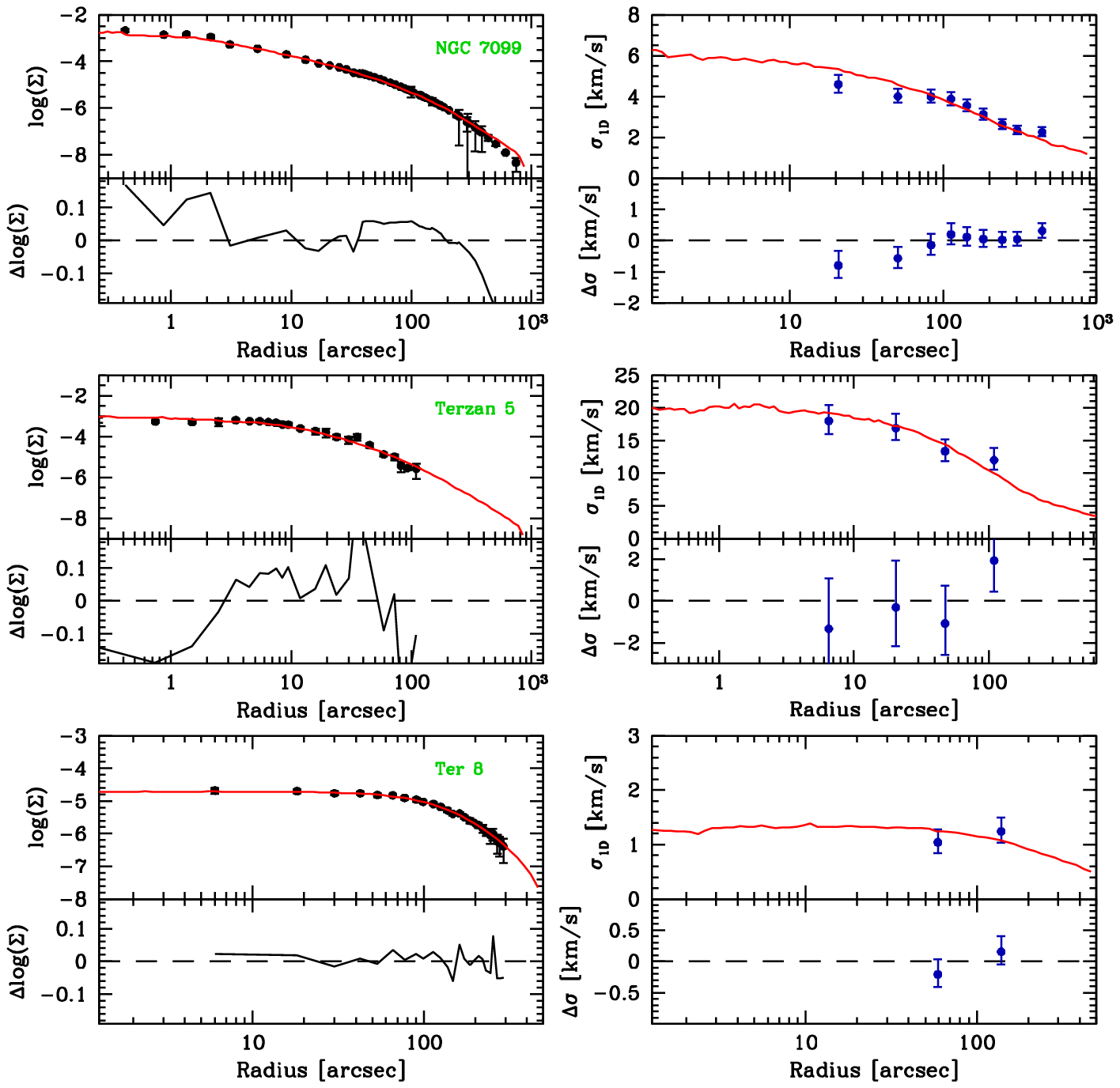}
\end{center}
\caption{Same as Fig. \ref{fig1a} for NGC 7099, Terzan 5 and Terzan 8.\hspace*{11cm}}
\label{fig15a}
\end{figure*}

\label{lastpage}
\end{document}